\documentclass[12pt,preprint]{aastex}
\usepackage{emulateapj5}

\newcommand{\zem}{{\ifmmode{z_{em}}\else{$z_{em}$}\fi}}
\newcommand{\zabs}{{\ifmmode{z_{abs}}\else{$z_{abs}$}\fi}}
\newcommand{\kms}{{\ifmmode{{\rm km~s}^{-1}}\else{km~s$^{-1}$}\fi}}
\newcommand{\cmm}{cm$^{-2}$}

\newcommand{\ergs}{{\ifmmode{{\rm erg~s}^{-1}}\else{erg~s$^{-1}$}\fi}}

\newcommand{\lognhi}{$\log N_{HI}$}
\newcommand{\delv}{$\Delta v$}

\newcommand{\hi}{{\rm H}~{\sc i}}

\newcommand{\lya}{Ly$\alpha$}
\newcommand{\lyb}{Ly$\beta$}
\newcommand{\lyg}{Ly$\gamma$}
\newcommand{\lyd}{Ly$\delta$}
\newcommand{\lye}{Ly$\epsilon$}
\newcounter{species}
\def\ion#1#2{\setcounter{species}{#2}#1$\;${\scriptsize\Roman{species}}\relax}

\newcommand{\noprint}[1]{}

\slugcomment{\today}
\slugcomment{A complete version can be obtained at http://www.astro.psu.edu/users/misawa/pub/Paper/40hires.ps.gz}
\shorttitle{HI Lines in 40 HIRES Quasars}
\shortauthors{Misawa et al.}

\begin{document}

\title{Spectroscopic Analysis of H I Absorption Line Systems in 40
HIRES quasars\altaffilmark{1}}

\footnotetext[1]{The data presented here were obtained at the
  W.M. Keck Observatory, which is operated as a scientific partnership
  among the California Institute of Technology, the University of
  California and the National Aeronautics and Space
  Administration. The Observatory was made possible by the generous
  financial support of the W. M. Keck Foundation.}

\author{Toru Misawa\altaffilmark{2},
        David Tytler\altaffilmark{3,4},
        Masanori Iye\altaffilmark{5,6},
        David Kirkman\altaffilmark{3,4},
        Nao Suzuki\altaffilmark{3,4},
        Dan Lubin\altaffilmark{3,4}, and
        Nobunari Kashikawa\altaffilmark{5,6}
}

\altaffiltext{2}{Department of Astronomy and Astrophysics,
  Pennsylvania State University, University Park, PA 16802}
\altaffiltext{3}{Center for Astrophysics and Space Sciences,
  University of California San Diego, MS 0424, La Jolla, CA
  92093-0424}
\altaffiltext{4}{Visiting Astronomer, W. M. Keck Observatory, which is
  a joint facility of the University of California, the California
  Institute of Technology, and NASA}
\altaffiltext{5}{National Astronomical Observatory, 2-21-1 Osawa,
  Mitaka, Tokyo 181-8588, Japan}
\altaffiltext{6}{Department of Astronomical Science, The Graduate
  University for Advanced Studies, 2-21-1 Osawa, Mitaka, Tokyo
  181-8588, Japan.}

\begin{abstract}

We list and analyze \hi\ absorption lines at redshifts 2 $<$ $z$ $<$ 4
with column density (12 $<$ $\log$($N_{\rm H\;I}$/[\cmm]) $<$ 19) in
40 high-resolutional (FWHM = 8.0~\kms) quasar spectra obtained with
the Keck$+$HIRES. We de-blend and fit all \hi\ lines within
1,000~\kms\ of 86 strong \hi\ lines whose column densities are
\lognhi\ $\geq$ 15~\cmm .  Unlike most prior studies, we use not only
\lya\ but also all visible higher Lyman series lines to improve the
fitting accuracy. This reveals components near to higher column
density systems that can not be seen in \lya . We list the Voigt
profile fits to the 1339 \hi\ components that we found.  We examined
physical properties of \hi\ lines after separating them into several
sub-samples according to their velocity separation from the quasars,
their redshift, column density and the S/N ratio of the spectrum.  We
found two interesting trends for lines with 12 $<$ $\log$($N_{\rm
H\;I}$/[\cmm]) $<$ 15 which are within 200 -- 1000~\kms\ of systems
with $\log$($N_{\rm H\;I}$/[\cmm]) $>$ 15.  First, their column
density distribution becomes steeper, meaning relatively fewer high
column density lines, at $z < 2.9$.  Second, their column density
distribution also becomes steeper and their line width becomes broader
by about 2--3~\kms\ when they are within 5,000~\kms\ of their quasar.

\end{abstract}


\section{Introduction}

Quasar absorption systems have historically been divided largely into
three physically distinct categories: (i) absorption systems with
strong metal lines arising in or near intervening galaxies, (ii) weak
\hi\ systems in the \lya\ forest that come from the intergalactic
medium, and (iii) intrinsic systems that are physically related to the
quasars, including associated and broad absorption line systems.

Metal absorption systems usually contain \hi\ lines with relatively
large column densities, including two sub-categories: damped Lyman
alpha (DLA) system and Lyman limit system (LLS) with \hi\ column
densities of $\log$($N_{\rm H\;I}$/[\cmm]) $>$ 20.2 and 17.16,
respectively. Deep imaging observations around quasars have provided
evidences that the metal absorption systems are often produced in
intervening galaxies. Galaxies have been detected that can explain
\ion{Mg}{2} absorption lines (e.g., Bergeron \& Boiss\'e 1991), and
\ion{C}{4} absorption lines (e.g., Chen, Lanzetta, \& Webb 2001).

On the other hand, nearly all \hi\ lines have smaller column densities
(\lognhi\ $\leq$ 15) than those associated found in gas that shows
strong metal lines.  The number of \hi\ lines per unit $z$ increases
with redshift (Peterson 1978; Weymann et al. 1998a; Bechtold 1994),
because the intergalactic medium is denser and less ionized at higher
$z$.  The \lya\ absorption lines are produced in intergalactic clouds
(e.g., Sargent et al. 1980; Melott 1980), which are the higher density
regions in the inter-galactic medium (IGM).  The \lya\ lines are
broadened by Hubble flow (e.g., Rauch 1998; Kim et al. 2002a) as well
as the Doppler broadening from the gas temperature.

Misawa et al. (2004) presented a study of \hi\ absorption lines seen
towards 40 quasars in spectra from the Keck HIRES spectrograph.  In a
departure from prior work, they considered the \hi\ lines without
considering the metal lines. They classified the \hi\ lines as either
high density lines [HDLs] which have or are near to strong \hi\ lines
that are probably related to galaxies, and low density lines [LDLs]
that are far from any strong \hi\ lines and are more likely to be far
from galaxies and hence in the IGM.

Following Misawa et al. (2004), we define HDLs as all \hi\ lines
within $\pm$ 200~\kms\ of a lines with $15$ $<$ \lognhi\ $<$
19~\cmm. We define LDLs as lines with 12 $<$ \lognhi\ $<$ 15 which are
within 200 -- 1000~\kms\ of a line with $15 < $ \lognhi\ $<$ 19 ~\cmm
.  This last velocity constraint is intended to make the LDL a
``control'' sample for the HDLs, where the two samples come from
similar redshifts and regions of the spectra with similar signal to
noise.
      
Misawa et al. (2004) discovered that the HDLs have smaller Doppler
parameters ($b$-values), for a given column density than the LDLs, and
they also found the same effect in a hydrodynamic simulation with 2.7
kpc cells. Misawa et al. (2004) suggested that the LDLs are cool or
shock-heated diffuse intergalactic gas, and that the HDLs are cooler
dense gas near to galaxies.

Misawa et al. (2004) fit all the accessible transitions in the \hi\
Lyman series to help de-blend \hi\ lines. Their main sample comprised
86 \hi\ absorption systems each with \lognhi\ $>$ 15 \cmm . They also
fit all \hi\ lines within $\pm$1,000~\kms\ of these \hi\ lines, to
give a total sample of 1339 \hi\ lines, including the 86 lines. This
is the only large sample in which multiple Lyman series lines are fit
together, although this method has been used on individual systems and
small samples (Songaila et al. 1994; Tytler, Fan, \& Burles 1996;
Wampler et al. 1996; Carswell et al. 1996; Songaila et al. 1997;
Burles \& Tytler 1998a,1998b, Burles, Kirkman, \& Tytler 1999; Kirkman
et al. 2000; O'Meara et al. 2001; Kirkman et al. 2003; Kim et
al. 2002b; Janknecht et al. 2006).

In this paper, we present measurements of the absorption lines that
Misawa et al. (2004) have analyzed. We give a detailed description of
each absorption system and we summarize new results. The paper is
organized as follows: In \S\ref{sec:data}, we give descriptions of the
data and the line fitting. The results of our statistical analyses are
presented in \S\ref{sec:results}. We discuss our results in
\S\ref{sec:discussion}, and summarize them in \S\ref{sec:summary}. In
the Appendix, we describe the properties and we give velocity plots
for each \hi\ system.  We use a cosmology in which $H_{0}=72$ \kms
Mpc$^{-1}$, $\Omega_{m}$ = 0.3, and $\Omega_{\Lambda}$ = 0.7.

\section{Spectra and Line Fitting\label{sec:data}}

The 40 quasars in our sample have either DLA systems or LLSs, and they
were observed as a part of a survey for measurements of the deuterium
to hydrogen (D/H) abundance ratio. The detailed description of the
absorption and data reduction are presented in Misawa et
al. (2004). We caution that our sample was biased in subtle ways by
the selection of LLS and DLAs that seemed more likely to show D, i.e.
those with simpler velocity structure.

We list the 40 quasars in Table~1. Column (1) is the quasar name, (2)
the emission redshift. Columns (3) and (4) are the optical magnitude
in the $V$ and $R$ bands. The lower and upper wavelength limits of the
spectra are presented in columns (5) and (6). Column (7) gives the S/N
ratio at the center of the spectrum.  Same data set was also used in
Misawa et al. (2007) in a study of the quasar intrinsic absorption
lines.

We will discuss only \hi\ lines with \lognhi\ $>$ 15 \cmm\ and other
\hi\ weaker lines within $\pm$1,000~\kms\ of these strong \hi\
lines. We selected this velocity range since it is enough to include
the conspicuous clustering of strong metal lines. Indeed such strong
metal lines are normally confined to an interval of $ <400$~\kms\ even
for DLA systems (Lu et al. 1996b).

Here we briefly review the line detection and fitting procedures that
we discussed in more detail in Misawa et al. (2004). We began
searching the literature for \hi\ lines with \lognhi\ $>$ 15,
including the DLA and LLS catalogues (Sargent, Steidel, \& Boksenberg
1989, hereafter SSB; Lanzetta 1991; Tytler 1982), and metal absorption
systems (P\'eroux et al. 2001; Storrie-Lombardi et al. 1996;
Petitjean, Rauch, \& Carswell 1994; Lu et al. 1993; Steidel \& Sargent
1992; Lanzetta et al. 1991; Barthel, Tytler, \& Thomson 1990; Steidel
1990a,b; Sargent, Boksenberg, \& Steidel 1988, hereafter SBS; SSB). We
also search for them ourselves. If more than one strong \hi\ line was
detected in a single 2000~\kms\ velocity window, we take the position
of the \hi\ line with the largest column density (hereafter the ``main
component'') as system center. We found 86 \hi\ systems with \lognhi\
$>$ 15, at 2.1 $<$ \zabs\ $<$ 4.0, in 31 of the 40 quasars. Figure~1
of Misawa et al. (2004) gives the velocity plot of one of these
systems, and below we give the rest.

We give parameters describing these 86 systems in Table~2. In
successive columns list (1) the name of the quasar; (2) the redshift
of the main component, that with the largest column density ; (3) the
\hi\ column density of the main component $N_{1}$; (4) $N_{2}$, the
second largest \hi\ column density within $\pm 1000$ \kms\ of the main
component; (5) the ratio of $N_{2}$ to $N_{1}$; (6) -- (9) the S/N
ratios at \lya, \lyb, \lye, and Ly10; (10) the number of lines in the
$\pm$ 1,000~\kms\ window; (11) the number of \hi\ lines classified as
HDLs (described later); (12) comments on the \hi\ system; (13)
references.  We will call this list sample S0 (Table~3).

When we were fitting the lines, we rejected narrow lines with Doppler
parameter of $b$ $<$ 4.8 \kms, which corresponds to the resolution of
our spectra. We also identify all lines with $b$ $<$ 15 \kms\ as
possible metal lines (called \ion{M}{1} in Tables) because \hi\ lines
with this narrow width are rare (e.g., Hu et al. 1995, hereafter H95;
Lu et al. 1996a, hereafter L96; Kirkman \& Tytler 1997a, hereafter
KT97). If there was more than one way to fit the lines, we chose the
fit with the fewest lines. If the model did not give good fits to all
the Lyman series lines, we adopted the model that best fit the lower
order lines where the SNR is best. For \hi\ lines with column
densities of \lognhi\ $\geq$ 16.6 the Lyman continuum optical depth is
$\tau$ $\geq$ 0.25. For these systems we checked if the residual flux
at the Lyman limit was consistent. Our fitting method could readily
overestimate the Doppler parameter but not the column density. Once
the fitting model is chosen, we used $\chi^{2}$ minimization in a code
written by David Kirkman, to get the best fit parameters of \hi\
column density (\lognhi), Doppler parameter ($b$), and absorption
redshift ($z$). The internal errors are typically
$\sigma$(\lognhi)=0.09~\cmm, $\sigma$($b$)=2.1~\kms , and
$\sigma$($z$)=2.5$\times$$10^{-5}$.

We prepared a sample S1 that is a sub-sample of S0 including only 973
\hi\ lines and 61 \hi\ systems with \lognhi\ $>$ 15. S1 excludes 25
systems with difficulties such as (i) poor fitting due to gaps in the
echelle formatted spectra, (ii) poor fitting due to strong DLA wings
(i.e., \lognhi\ $>$ 19), (iii) close proximity in redshift to the
background quasars (i.e., within 1,000~\kms\ of the emission
redshift), and (iv) overlapping with other \hi\ systems.  The S/N
ratios of the spectra are at least S/N $\simeq$ 11 per 2.1~\kms\ pixel
and the mean value is S/N $\simeq$ 47 for \lya\ lines. Among these 61
\hi\ systems, three systems may be physically associated with the
quasars based on the partial coverage analysis for the corresponding
metal absorption lines (Misawa et al. 2007). However, we keep these
systems in S1 sample, because we still cannot reject the idea that
they are intervening systems.

We give detailed descriptions of all the lines that we fit in the
Appendix. We also give velocity plots of the first five Lyman
transitions, \lya, \lyb, \lyg, \lyd, and \lye .

\section{RESULTS\label{sec:results}}

We investigate the properties of line parameters such as the column
density, Doppler parameter, and the clustering properties of the \hi\
lines. Since this sample contains not only \hi\ lines originating in
the intergalactic diffuse gas clouds (i.e., LDLs), but also \hi\ lines
produced by intervening galaxies (i.e., HDLs), we also attempted to
separate \hi\ lines into HDLs and LDLs based on the clustering trend
(Misawa et a. 2004).

Our analysis is similar to that of previous studies (e.g., H95; L96;
KT97), but with three key differences: (i) earlier studies used all
\hi\ lines detected in the quasar spectra, whereas we use only \hi\
lines within $\pm$ 1,000~\kms\ of the main components with \lognhi\
$\geq$ 15, (ii) our sample contains a number of strong lines (\lognhi\
$\geq$ 15) in addition to weak \hi\ lines (\lognhi\ $<$ 15), and (iii)
our sample covers a wide redshift range: 2.0 $\leq$ \zabs\ $\leq$
4.0. The redshift distributions of the 86 and 61 \hi\ systems in
samples S0 and S1 are shown in Figure~1.

\subsection{Sub-Samples for the Statistical Analysis}

For further investigation, we prepared several sub-samples as
follow. It is known that the comoving number densities of
low-ionization lines, such as \hi\ lines, decreases in the vicinity of
quasars (Carswell et al. 1982; Murdoch et al. 1986; Tytler 1987). This
trend is known as the ``proximity effect'', and is probably caused by
the enhanced UV flux from the quasar towards which the absorption is
seen.  We separate the 61 \hi\ systems (sample S1) into sub-samples
S2a (the velocity difference from the quasar, $\Delta v$ $>$
5,000~\kms) and S2b ($\Delta v$ $<$ 5,000~\kms). We have already
removed from S1 all \hi\ systems within 1,000~\kms\ of the quasars to
avoid \hi\ systems from the quasar host galaxies.

H95 emphasized that the line detection limit is almost wholly
determined by the line blending (or blanketing), and not by the S/N
ratio of the spectrum. In order to confirm whether the distribution of
line parameters is affected by the quality of the spectrum, we made
two overlapping samples from S1 using the S/N ratio of each spectrum
in the \lya\ region: S3a (S/N $\geq$ 40), and S3b (S/N $\geq$
70). These sub-samples include 34 ($\sim$ 60\%) and 17 ($\sim$ 30\%)
of the 61 \hi\ systems of the sample S1.

Sample S1 covers a broader range of redshifts, $2.0 < z < 4.0$, when
compared with previous studies: $2.55 < z < 3.19$ for H95, $3.43 < z <
4.13$ for L96, and $2.43 < z < 3.05$ for KT97. When we investigate the
redshift evolution of \hi\ absorbers, we also divided S1 into two
sub-samples; S4a ($z < 2.9$) and S4b ($z \geq 2.9$). Here the two
sub-samples have nearly the same number of \hi\ systems.

Finally, we made sub-samples according to the column densities of \hi\
lines, as the distributional trends of strong and weak \hi\ lines are
very different (see Figure~2 in Misawa et al. 2004); the \hi\ lines
with relatively large column densities tend to cluster around the main
components, while the number of weak \hi\ lines decreases near the
center of \hi\ systems because of line blanketing. Since one of our
interests is to determine the boundary value of column density between
HDLs and LDLs (although other parameters may be necessary to separate
them), we separate the 973 \hi\ lines into eight sub-samples according
to their column densities. We use boundary values of $\log N$ = 13,
14, 15, and 16, where sub-sample S5$_{ab}$ contains \hi\ lines whose
column densities are $\log$($N_{\rm H\;I}$/[\cmm]) values of a to b.

In Table~3 we summarize these 16 sub-samples. Here we should emphasize
that S5$_{1213}$, S5$_{1214}$, S5$_{1215}$, S5$_{1216}$, S5$_{1319}$,
S5$_{1419}$, S5$_{1519}$, and S5$_{1619}$, are separated by the
properties of lines, while S2a, S2b, S3a, S3b, S4a, and S4b are
separated by the unit of system.

\subsection{Physical Properties of \hi\ Absorbers}

For each sub-sample prepared in the last section, we perform
statistical analysis, including analysis of the column density
distribution, Doppler parameter distribution, and line clustering
properties. The samples used here contain both HDLs and LDLs.

\subsubsection{Column Density Distribution}

The column density distribution of \hi\ lines, $dn(N)/dN$, are usually
fit with a power law (Carswell et al. 1984, Tytler 1987, Petitjean et
al. 1993; H95),
\begin{eqnarray}
\log\left(\frac{{\rm d}n(N)}{{\rm d}N}\right) = - \beta
\hspace{1mm} \log N + A,
\label{eqn:1}
\end{eqnarray}
where the index $\beta$ was estimated to be 1.46 (H95), 1.55 (L96),
and 1.5 (KT97) with only weaker \hi\ lines with \lognhi\ = 12 --
14.5. Janknecht et al. (2006) found $\beta$ = 1.60$\pm$0.03 at lower
redshift $z$ $<$ 1.9. The column density distributions of our seven
sub-samples (S1, S2a, S2b, S3a, S3b, S4a and S4b) per unit redshift
and unit column density are analyzed. The plot in Figure~2 is the
result for sub-sample S1. Since the turn-over of the distribution
around \lognhi\ = 12.5 is probably due to line blending and/or
blanketing as described later in \S\ 4, we fit the column density
distributions only for \lognhi\ $>$ 13. The best-fitting parameters,
$\beta$ and $A$, as well as the redshift bandpass, $\Delta z$, of each
sub-sample are summarized in Table~4, along with the past results from
KT97 and Petitjean et al. (1993).

The indices that we find, $\beta$ = 1.40$\pm$0.03, are slightly
smaller than the value in the past results, $\beta$ = 1.46 --1.55,
which means that our sample favors strong \hi\ lines. But we expect
this type of trend.  Our samples contain not only LDLs but also HDLs,
and since we cover only the velocity regions within $\pm$ 1,000~\kms\
of the main components, we have a strong excess of strong \hi\
lines. We see the column density distribution does not change with the
velocity distance from the quasars (S2a and S2b) or with the S/N ratio
(S3a and S3b), but it is weakly affected by redshift (S4a and S4b). We
also applied the Kolmogorov-Smirnov (K-S) test to the sub-samples. The
results in Table~5 show that we can not rule out the hypothesis that
they are random samplings from the same population.

\subsubsection{Doppler Parameter Distribution}

The distribution of the Doppler parameter of \hi\ lines have been
approximately given by the truncated Gaussian distribution (H95; L96),
\begin{equation} \frac{{\rm d}n(b)}{{\rm d}b} = \left\{
\begin{array}{ll} A \;
\exp \left[ \frac{-(b-b_{0})^{2}}{2 \sigma_{b}^{2}} \right] & b \geq
b_{min}\\
0 & b < b_{min}
\end{array}
\right.
\label{eqn:2}
\end{equation}
where $b_{0}$ and $\sigma_{b}$ are the mean and the dispersion of $b$
distribution and $b_{min}$ is the minimum $b$ value for an \hi\ line.
There are two origins of line broadening: thermal broadening
($b_{T}$), and micro-turbulence ($b_{tur}$). The total amount of
broadening is given by $b = \sqrt{b_{T}^{2} + b_{tur}^{2}}$.

In order to determine the correct Doppler parameter, we have to
individually resolve and fit \hi\ lines using Voigt profiles. However,
most of weak \hi\ lines disappear in the observed spectrum due to line
blending and blanketing, especially near strong lines. To derive the
intrinsic (as opposed to observed) distribution of the Doppler
parameter, previous authors (i.e., H95; L96; KT97) have chosen
artificial \hi\ lines (as input data) with distributions characterized
by a Gaussian, and used them for comparison with the observed
distributions. KT97 noted that the $b$-value distribution of lines
seen in simulated spectra resembles the distribution of the input
data, except for two differences: (i) an excess of lines with large
$b$-values is seen in the recovered data, compared to the input data,
which is probably produced by line blending, and (ii) lines with very
small $b$-values ($b$ $<$ 15~\kms) are found, which are probably data
defects or noise, as they are not present in the input
data. Nonetheless, the $b$-value distribution of the input and
recovered data resemble each other between $b$ = 20~\kms\ and $b$ =
60~\kms.

We would ideally like to determine the real distribution of Doppler
parameters; however, the only way to do this is to perform
simulations, and compare the recovered Doppler parameter distribution
with the observed distribution. Such simulations are expensive to
perform, so in this work we simply compared the distribution in our
sample with the past results of H95 and L96 at $b$ = 20 -- 60~\kms. We
analyzed the observed distributions of Doppler parameters for 15
sub-samples, and compare them with the results in H95 ($b_{0}$ =
28~\kms, $\sigma_{b}$ = 10~\kms, and $b_{min}$ = 20~\kms) and L96
($b_{0}$ = 23~\kms, $\sigma_{b}$ = 8~\kms, and $b_{min}$ = 15~\kms),
where the parameters are the inputs used for artificial spectra that
reproduce the observed distribution.

We see an excess of \hi\ lines with large $b$-values of $b$ $>$
50~\kms\ in all of the sub-samples, while we have no lines with $b$
$\leq$ 15~\kms\ because we decided to classify them into metal
lines. All sub-samples except for $S_{1213}$ have relatively large
$b$-values, and their distributions are closer to H95's distribution
than L96's. In contrast, the distribution of $S_{1213}$, containing
only \hi\ lines with small column densities, \lognhi\ $<$ 13,
resembles L96's distribution.  We plot in Figure~3 the Doppler
parameter distributions of sub-samples S1 and S$_{1213}$.  We also
applied a K-S test to the sub-samples (S2a, S2b, S3a, S3b, S4a, and
S4b). The results are listed in Table~6. The probability, that the
distributions of sub-samples S2a ($\Delta v$(\zem $-$ \zabs) $>$
5000~\kms) and S2b ($\Delta v$(\zem $-$ \zabs) $\leq$ 5000~\kms) were
drawn from the same parent population, is very small, $\sim$2.4~\%.
This result could suggest that the Doppler parameters of \hi\ lines
within 5000~\kms\ of quasars are affected by UV flux from the quasars.

\subsection{HDLs and LDLs}

In the previous section, we carried out statistical analysis using
sub-samples containing both HDLs and LDLs together.  Here, we repeat
these tests on the two samples separately.

Metal absorption lines seen in DLA systems or LLSs are strongly
clustered within several hundred \kms , which implies their
relationship to galaxies. In simulations Dav\'e et al. (1999) also
noted that galaxies tend to lie near the dense regions that are
responsible for strong \hi\ lines. On the other hand, for weak \hi\
lines, no strong clustering is seen (e.g., Rauch et al. 1992; L96;
KT97), although some studies found only weak clustering trends (e.g.,
Webb 1987; H95; Cristiani et al. 1997).

As Misawa et al. (2004) found, lines with \lognhi\ $\geq$ 15 show
strong clustering trends at $\Delta v$ $<$ 200~\kms, while lines with
lower column densities cluster weakly at $\Delta v$ $<$ 100~\kms\
(Figure~4).

Misawa et al. (2004) defined HDLs as \hi\ lines with $15$ $<$ \lognhi\
$<$ 19 and other weaker \hi\ lines within $\pm$ 200~\kms\ of those
stronger \hi\ lines. They then defined the LDLs as all other lines
with 12 $<$ \lognhi\ $<$ 15. They chose \lognhi\ = 15 ~\cmm\ in the
definition because the two point correlation was largest for a
sub-sample of \hi\ lines with 15 $<$ \lognhi\ $<$ 19. We list the
number of HDLs and LDLs in the sub-samples in Table~7.

\subsubsection{Column Density Distribution}

In Table~8 we give the parameters that describe the column density
distributions of HDLs and LDLs for five sub-samples (S1, S2a, S2b,
S4a, and S4b). The most obvious result is that the HDLs have a smaller
index than the LDLs. We see the same result in Figure~3 of Petitjean
et al. (1993), and hence we now confirm this with the first large
sample to consider the sub-components of the HDLs.

The distributions of LDLs in sample S1, S2a, and S4b are almost
consistent with the previous result in KT97. On the other hand, the
power law indices for the LDLs of S2b ($\Delta v$ $\leq$ 5000~\kms)
and S4a ($z$ $<$ 2.9), $\beta$ = 1.90$\pm$0.16 and 1.71$\pm$0.06, are
larger than the values for the other sub-samples, $\beta \sim$ 1.52,
which means that LDLs at lower redshift or near the quasars tend to
have lower column densities compared with those at higher redshift or
far from the quasars.  The change in the column density distribution
near to the quasars may be just a consequence of the enhanced UV
radiation.

\subsubsection{Doppler Parameter Distribution}

The Doppler parameter distributions of HDLs and LDLs for sub-samples
(S1, S2a, S2b, S4a, and S4b) are also investigated, and the results of
K-S tests applied to them are summarized in Table~9. The only
remarkable result is that the probability, that the Doppler parameter
distributions of LDLs at $\Delta v$ $>$ 5,000~\kms\ (S2a) and $\Delta
v$ $\leq$ 5,000~\kms\ (S2b) from the quasars were drawn from the same
parent population, is very small, $\sim$2.0\%.  We shows these two
distributions in Figure~5.  In Figure~6 we see that the cumulative
distribution for the line $b$-values rises more slowly for the LDLs
near to the quasars (at $\Delta v$ $\leq$ 5000~\kms), which means that
these lines near to the quasars are broader by about 2--3~\kms.  For
other pairs of the sub-samples, we could not rule out the hypothesis
that their parent populations are same.

\section{Discussion\label{sec:discussion}}

In this section, we discuss our results, especially the fact that the
column density distribution is changing with the redshift and the
velocity distance from the quasars. We also compare our results to
those at lower redshift ($z$ $<$ 0.4) from the literature. After that,
we also briefly discuss the completeness of \hi\ lines in our 40
spectra.

\subsection{Redshift Evolution of \hi\ Absorbers}

In \S\ 3, we prepared two sub-samples, S4a and S4b, to compare the
physical properties of \hi\ lines at lower redshift (\zabs\ $<$ 2.9)
and at higher redshift (\zabs\ $\geq$ 2.9). We do not see a change in
the column density distribution in the sample as a whole, but once
they are separated into HDLs and LDLs, we notice that the index of the
column density distribution of LDLs at \zabs\ $<$ 2.9 ($\beta$ =
1.71$\pm$0.06) is clearly different from that of LDLs at \zabs\ $\geq$
2.9 ($\beta$ = 1.52$\pm$0.09). On the other hand, there was no
redshift evolution for HDLs. This trend, shown in Figure~7, means that
there is a deficit of relatively stronger LDLs (i.e., \lognhi\ $\geq$
14.5) at lower redshift. One of the possible explanations is that at
lower redshift, more \hi\ lines with the column densities just below
\lognhi\ = 15 (i.e., relatively strong LDLs) might be associated with
HDLs. In other words, stronger (i.e., \lognhi\ = 14.5 -- 15) LDLs get
into within 200~\kms\ of the nearest HDLs, and would be classified
into HDLs, which is consistent with the trend expected in the
hierarchical clustering model (Figure~8).

As for the Doppler parameter distribution, we did not find any
remarkable redshift evolutions in neither HDLs nor LDLs. L96 claimed
that there is a redshift evolution of the Doppler parameter between
$z$ = 2.8 and 3.7; the mean value of Doppler parameter at \zabs\ = 3.7
($b_{0}$ = 23~\kms; L96) is smaller than the value at \zabs\ = 2.8
($b_{0}$ = 28~\kms; H95). The corresponding value in KT97 ($b_{0} =
23$ \kms) is, however, different from the result in H95. The
difference may be due to the different line fitting procedure used in
these studies; L96 and KT97 used the VPFIT software, while H95 used
different software. Especially important is how the authors chose to
treat blended lines. The difference could be related to the difference
of the spectrum resolutions; R=45000 (L96; KT97) and R=36000 (H95).
Janknecht et al. (2006) did not detect any evolution on the Doppler
parameter at $z$ = 0.5 -- 1.9. Our results, which are based on the
data set taken with one observational configuration and fit using the
same procedure, suggests that the Doppler parameter distribution of
\hi\ clouds does not evolve with redshift at $z$ = 2 -- 4.

\subsection{Proximity Effect near Quasars}

It has long been noted that the number of \lya\ lines decreases near
to the redshift of the quasars (Carswell et al. 1982; Murdoch et
al. 1986; Tytler 1987). This phenomenon is related to the local excess
of UV flux from the quasars. The proximity effect has been used to
evaluate the intensity of the background UV flux. Bajtlik et
al. (1988) first measured the mean intensity of the background UV
intensity, $J_{\nu}$ = $10^{-21.0\pm0.5}$ (erg s$^{-1}$ cm$^{-2}$
Hz$^{-1}$ str$^{-1}$) at the Lyman limit at 1.7 $<$ $z$ $<$ 3.8, by
estimating the distance from the quasar at which the quasar flux is
equal to the background UV flux. The typical radius is $\sim$ 5~Mpc in
physical scale that corresponds to the velocity shift of $\Delta v$
$\sim$ 4,000~\kms\ from the quasars. L96 also evaluated the background
UV intensity to be $J_{\nu}$ = $2\times10^{-22}$ (erg s$^{-1}$
cm$^{-2}$ Hz$^{-1}$ str$^{-1}$) at $z$ $\sim$ 4.1 in the spectrum of
Q0000-26.

We see two differences in LDLs within 5,000~\kms\ of the quasars,
compared to those far from the quasars ($\Delta v$ $>$ 5,000~\kms). We
see fewer strong LDLs leading to a large index for the column density
power law, $\beta$ = 1.90$\pm$0.16 (Figure~9). We also see that the
distribution of Doppler parameter is different from that of \hi\ lines
far from the quasars at 98.8~\% confidence level. The lines near to
the quasar apparently tend to have broader lines (Figures~5 and 6),
although this is a tentative result because we consider few lines near
to the quasars.

These results could be accounted for by assuming a two-phase
structure: outer cold low-density regions and inner hot high-density
regions in which temperature is determined by the competition between
photoionization heating and adiabatic cooling. When gas is near to the
quasars, the outer regions become too highly ionized to show much
\ion{H}{1}, and only the inner hot regions would be observed in
\ion{H}{1}, which would increase the mean value of the Doppler
parameter. The increased ionization also decreases the total column
densities of \ion{H}{1} gas compared with gas far from the quasars
(Figure~10).  As reported in the past observations (e.g., Kim et
al. 2001; Misawa et al. 2004), \lognhi\ and $b$(\hi) have a positive
correlation for \lognhi\ $<$ 15.  This correlations was also
reproduced by hydrodynamical simulations (e.g., Zhang et al. 1997;
Misawa et al. 2004). These results suggest that high density regions
tend to have larger Doppler parameters, if the absorbers are not
optically thick.  Dav\'e et al. (1999) also presented an interesting
plot in their Figure~11 that supported there existed three kinds of
phases for \hi\ absorbers (diffuse, shocked, and condensed phases).
Among them, the diffuse phase whose volume densities are small (i.e.,
it corresponds to LDLs in our paper) has a positive correlation
between \lognhi\ and $b$. On the other hand, an anti-correlation
between \lognhi\ and $b$ is seen only for the condensed phase with
high volume density that is probably associated with galaxies.  The
shocked phase, probably consisting of shock-heated gas in galaxies,
does not show any remarkable correlations between them.  Thus, if we
assume all LDLs in our sample arise in the diffuse phase absorbers,
our scenario above could reproduce the difference between sub-samples
S2a and S2b.

\subsection{Comparison to \hi\ Absorbers at Lower Redshift}

The number density evolution of \hi\ absorbers (i.e., $dN/dz$
$\propto$ $(1+z)^{-\gamma}$) has been known to slow dramatically at
$z$ $\sim$ 1.6, from a high-$z$ rapid evolution with $\gamma$ of
1.85$\pm$0.27 (Bechtold 1994) to a low-$z$ slow evolution with
$\gamma$ of 0.16$\pm$0.16 (Weymann et al. 1998b). This trend is
suggested to be due to the decline in the extragalactic background
radiation using hydrodynamic cosmological simulations (e.g., Theuns,
Leonard, \& Efstathiou 1998). Thus, a comparison of \hi\ absorbers at
high-$z$ and local universe is another interesting topic.

In \S~3, we found that the column density distribution of LDLs at $z$
$<$ 2.9 ($\beta$ = 1.71$\pm$0.06) is steeper than that at $z$ $>$ 2.9
($\beta$ = 1.52$\pm$0.09). We proposed this trend could be due to the
hierarchical clustering. If the assembly of structure in the IGM
indeed dominates the column density distribution, we would expect to
find a steeper column density distributions at lower redshift as
proposed in \S~4.1.

Using the {\it Hubble Space Telescope} ({\it HST}) and the {\it Far
Ultraviolet Spectroscopic Explorer} ({\it FUSE}), Penton, Stocke, \&
Shull (2004) and Lehner et al. (2007) estimated power-law indices to
be $\beta$ of 1.65$\pm$0.07 at 12.3 $\leq$ \lognhi\ $\leq$ 14.5 and
1.76$\pm$0.06 at 13.2 $\leq$ \lognhi\ $\leq$ 16.5 at $z$ $<$ 0.4,
respectively. On the other hand, Dav\'e \& Tripp (2001) found a
flatter distribution ($\beta$ = 2.04$\pm$0.23) at $z$ $<$ 0.3. The
latter steeper distribution was also reproduced by hydrodynamical
simulations (e.g., Theuns et al. 1998). If we accept the steeper
result, the column density distribution would continue to be steeper
as going to the lower redshift, which supports our idea that the
hierarchical clustering could play a main role of the evolution seen
in the column density distribution, although extragalactic radiation
would contribute to play a role.

Absorption line width is another parameter that is still in argument
whether it would evolve with redshift or not, as mentioned in \S~4.1.
While most of the space-based ultraviolet observations could not
measure line widths by model fittings because of the lacks of spectral
resolutions, Lehner et al. (2007) for the first time measured Doppler
parameters of \hi\ absorption lines accurately at lower redshift ($z$
$<$ 0.4), and investigate their distributional trend. By comparing to
the results at higher redshift, Lehner et al. (2007) discovered that
Doppler parameters are monotonously increasing from $z$ = 3.1 to
$\sim$0. Such a trend was not confirmed in past papers (e.g.,
Janknecht et al. 2006).  The fraction of the broad \lya\ absorbers
(BLA; $b$ $\geq$ 40~\kms) is also confirmed to increase by a factor of
$\sim$3 from $z$ $\sim$ 3 to 0 (Lehner et al. 2007). Here, $b$ =
40~\kms\ corresponds to gas temperature of $T_{gas}$ $\sim$
$10^{5}$~K, which is a border between the cool photoionized absorbers
and the highly ionized warm-hot absorbers.  These results suggests
that a large fraction of \hi\ absorbers at very low redshift (i.e.,
$z$ $<$ 0.4) are hotter and/or more kinematically disturbed than at
higher redshift (i.e., $z$ $>$ 2.0).

In our sample,we do not see any clear difference of mean/median
Doppler parameter at $z$ $\geq$ 2.9 ($b_{mean}$ = 31.0$\pm$10.0,
$b_{med}$ = 28.1) and $z$ $<$ 2.9 ($b_{mean}$ = 32.0$\pm$10.9,
$b_{med}$ = 29.6). Neither HDLs nor LDLs shows any evolutional
trends. These negative results could be because with our optical data
we covered only higher redshift regions than $z$ $\sim$ 1.6, at which
$dN/dz$ evolution dramatically changed. Similarly, the fraction of the
BLA ($f_{BLA}$ = 0.182 at $z$ $\geq$ 2.9 and 0.196 at $z$ $<$ 2.9)
shows an only marginal hint to the evolution. However, these fractions
are consistent to the result from KT97 ($f_{BLA}$ = 0.179; Lehner et
al. 2007) at 2.43 $<$ $z$ $<$ 3.05 that is similar redshift coverage
as our sample.  Thus, Doppler parameter could increase as going to the
lower redshift, but such a trend would be remarkable only if we trace
its distribution at very low redshift (at $z$ $<$ 0.4) and compare it
to that at much higher redshift (at $z$ $>$ 2).

As for the clustering trend of \hi\ absorption lines, we see a very
similar property at low and high redshift regions. As presented in
Figure~4, we found a strong clustering trend within \delv\ of
200~\kms\ for \hi\ lines with \lognhi\ between 15 and 19, while only a
weak correlation is seen for weaker \hi\ lines within \delv\ of
100~\kms. Penton et al. (2004) presented very similar results:
5$\sigma$ (7.2$\sigma$) excess within \delv\ of 190~\kms\ (260~\kms)
and only stronger \hi\ lines contribute to this clustering. Penton,
Stock, \& Shull (2002) proposed such clustering trends within several
hundreds of \kms\ are due to clusters of galaxies. There could exist
similar kinematical structures both at high-$z$ and in local universe.

\subsection{Completeness of \hi\ Line Sample}

For a statistical analysis, especially the number density analysis,
the completeness of the \hi\ line detection is influenced by the
detection limit of absorption lines (e.g., equivalent width or column
density). In this study, we have used the \hi\ lines detected in the
40 HIRES spectra that have various S/N ratios. The strong line sample
will have subtle biases arising from the selection of the quasars
because they were once thought to be good targets for the detection of
deuterium. For example, we avoided quasars with no LLSs, and we
avoided LLSs with previously known complex velocity
structure. Nevertheless, we confirmed that our sample is almost
complete for weak lines in the following way.

The minimum detectable equivalent width in the observed-frame,
$W_{min}$, can be estimated using the following relation,
\begin{equation}
 U = \frac{W_{min}N_{C}}{\sigma(W_{min}N_{C})}
   = \frac{W_{min}(S/N)}{(M_{L}^{2}M_{C}^{-1}+M_{L}-W_{min})^{1/2}},
\label{eqn:3}
\end{equation}
where $M_{L}$ and $M_{C}$ are the numbers of pixels over which the
equivalent width and the continuum level ($N_{C}$) are determined
(Young et al. 1979; Tytler et al. 1987). The value of ($S/N$) is the
S/N ratio per pixel. When we set $U \simeq W/\sigma(W)$ is 4 (i.e.,
4$\sigma$ detection), the eqn.(\ref{eqn:3}) can be solved to give,
\begin{eqnarray}
 W_{min} & = & (S/N)^{-2}\{[64+16(S/N)^{2} \times \nonumber \\
         &   & (M_{L}+M_{L}^{2}/M_{C})]^{1/2}-8\} \times 
                \Delta \lambda \hspace{2mm} (\rm{\AA}),
\label{eqn:4}
\end{eqnarray}
where $\Delta \lambda$ is the wavelength width per pixel in angstroms
(Misawa et al. 2002). Here, we set $M_{L}$ for 2.5 times FWHM of each
line, and $M_{C}$ for full width of each echelle order. Once the
minimum rest-frame equivalent width, $W_{rest} [=W_{min}/(1+z)]$, has
been evaluated, it can be converted to the minimum column density by
choosing a specific Doppler parameter; the result is insensitive to
the choice on the linear part of the curve of growth.  Among the 86
\hi\ systems in our data sample, the \hi\ system at \zabs\ = 2.940 in
the spectrum of Q0249-2212 is located in the region with the lowest
S/N ratio (i.e., $S/N$ $\sim$ 11). This corresponds to a 4$\sigma$
detection limit of \lognhi\ $\sim$ 12.3 for an isolated \lya\ line
with any Doppler parameter seen in our sample ($b$ = 15 -- 80~\kms).
Thus, our sample is complete for \hi\ lines with \lognhi\ $>$
12.3. Therefore, the bend in the column density distribution near
\lognhi\ $\sim$ 13 is probably due to the line blending and
blanketing.

\section{Summary\label{sec:summary}}

We present 40 high-resolutional (FWHM = 8.0~\kms) spectra obtained
with Keck+HIRES. Over the wide column density range (12 $<$ \lognhi\
$<$ 19), we fit \hi\ lines by Voigt profiles using not only \lya\ line
but also higher Lyman series lines such as \lyb\ and \lyg\ up to Lyman
limit when possible. To investigate the detailed line properties, we
made several sub-samples that are separated according to the distance
from the quasar, redshift, the column density, and the S/N ratio of
the spectrum. We also classify them into HDLs (lines arising in or
near to intervening galaxies) and LDLs (lines not obviously near to
galaxies and hence more likely to be from the intergalactic diffuse
gas), based on the clustering properties. The main results are
summarized below:

\begin{enumerate}

\item
We present a database of \hi\ absorption lines with a wide column
density range (i.e., \lognhi\ = 12--19) from a wide redshift range
(i.e., $z$ = 2--4).

\item
Our data sample is complete at \lognhi\ $\geq$ 12.3 with 4$\sigma$
line detection. The turnover at \lognhi\ $<$ 13 seen in the \lognhi\
distribution is not due to a quality of our spectra but due to the
line blending and blanketing.

\item
The power-law indices of the column density distribution of LDLs shows
evolution with redshift, from $\beta$ = 1.52$\pm$0.09 at $z$ $\geq$
2.9 to $\beta$ = 1.71$\pm$0.06 at $z$ $<$ 2.9. This trend could be
related to the hierarchical clustering in cosmological timescale. No
evolution is seen for HDLs.

\item
Within 5,000~\kms\ of the quasars, the power-law index of the column
density distribution for LDLs ($\beta$ = 1.90$\pm$0.16) is larger than
those far from the quasars ($\beta$ = 1.53$\pm$0.05). We also found a
hint (Figure~6) that the Doppler parameters are larger near the
quasars.  These results could be due to the UV flux excess from the
quasars. We do not see any similar trend for the HDLs.

\item
We suggest that HDLs and LDLs are produced by physically different
phases or absorbers, because they have four key differences seen in
(i) clustering property, (ii) redshift evolution, (iii) Proximity
effect, and (iv) \lognhi\ -- $b_{min}$ relation (see Misawa et
al. 2004).

\end{enumerate}

\acknowledgments We acknowledge support from NASA under grant
NAG5-6399, NAG5-10817, NNG04GE73G and by the National Science
Foundation under grant AST 04-07138. This work was also in part
supported by JSPS. The UCSD team were supported in part by NSF grant
AST 0507717 and by NASA grant NAG5-13113. We also thank the anonymous
referee for very useful comments and suggestions.


\clearpage



\begin{deluxetable}{llccccc}
\tabletypesize{\scriptsize}
\tablecaption{Keck HIRES Spectra of 40 quasars \label{tab:1}}
\tablewidth{0pt}
\tablehead{
\colhead{(1)} &
\colhead{(2)} &
\colhead{(3)} &
\colhead{(4)} &
\colhead{(5)} &
\colhead{(6)} &
\colhead{(7)} \\
\colhead{quasar$$$^{a}$} &
\colhead{\zem} &
\colhead{$m_V$$^{b}$} &
\colhead{$m_R$$^{c}$} &
\colhead{$\lambda_{min}$$^{d}$} &
\colhead{$\lambda_{max}$$^{e}$} &
\colhead{S/N$^f$} \\
\colhead{} &
\colhead{} &
\colhead{} &
\colhead{} &
\colhead{(\AA)} &
\colhead{(\AA)} &
\colhead{} \\
}
\startdata
Q0004+1711     &  2.890 & 18.70 &       & 3510 & 5030 & 11.9 \\
Q0014+8118     &  3.387 &       & 16.1  & 3650 & 6080 & 48.8 \\
Q0054-2824     &  3.616 &       & 17.8  & 4090 & 6510 & 18.2 \\
Q0119+1432     &  2.870 & 17.4  &       & 3200 & 4720 & 23.7 \\
HE0130-4021    &  3.030 & 17.02 &       & 3630 & 6070 & 52.5 \\
Q0241-0146     &  4.040 & 18.20 &       & 4490 & 6900 &  7.5 \\
Q0249-2212     &  3.197 & 17.70 &       & 3500 & 5020 & 11.0 \\
HE0322-3213    &  3.302 & 17.80 &       & 3830 & 5350 & 12.8 \\
Q0336-0143     &  3.197 &       & 18.8  & 3940 & 6390 & 12.7 \\
Q0450-1310     &  2.300 & 16.50 &       & 3390 & 4910 & 17.1 \\
Q0636+6801     &  3.178 &       & 16.9  & 3560 & 6520 & 53.4 \\
Q0642+4454     &  3.408 &       & 18.4  & 3930 & 6380 & 19.0 \\
HS0757+5218    &  3.240 & 17.3  &       & 3590 & 5120 & 21.5 \\
Q0805+0441     &  2.880 & 18.16 &       & 3800 & 6190 & 15.7 \\
Q0831+1248     &  2.734 & 18.10 &       & 3790 & 6190 & 29.2 \\
HE0940-1050    &  3.080 & 16.90 &       & 3610 & 6030 & 35.7 \\
Q1009+2956     &  2.644 & 16.40 &       & 3090 & 4620 & 48.6 \\
Q1017+1055     &  3.156 &       & 17.2  & 3890 & 6300 & 44.8 \\
Q1055+4611     &  4.118 & 17.70 &       & 4450 & 6900 & 40.3 \\
HS1103+6416    &  2.191 & 15.42 &       & 3180 & 5790 & 78.1 \\
Q1107+4847     &  3.000 & 16.60 &       & 3730 & 6170 & 51.8 \\
Q1157+3143     &  2.992 & 17.00 &       & 3790 & 6190 & 35.9 \\
Q1208+1011$^g$ &  3.803 &       & 17.2  & 3730 & 6170 & 21.8 \\
Q1244+1129     &  2.960 & 17.70 &       & 3370 & 4880 &  9.9 \\
Q1251+3644     &  2.988 & 19.00 &       & 3790 & 6190 & 32.5 \\
Q1330+0108     &  3.510 &       & 18.56 & 4030 & 6450 &  8.8 \\
Q1334-0033     &  2.801 & 17.30 &       & 3730 & 6170 & 24.7 \\
Q1337+2832     &  2.537 & 19.30 &       & 3170 & 4710 & 31.1 \\
Q1422+2309$^h$ &  3.611 &       & 15.3  & 3740 & 6180 & 136  \\
Q1425+6039     &  3.165 &       & 16.0  & 3730 & 6170 & 43.5 \\
Q1442+2931     &  2.670 & 16.20 &       & 3740 & 6180 & 29.4 \\
Q1526+6701     &  3.020 & 17.20 &       & 3460 & 4980 &  9.7 \\
Q1548+0917     &  2.749 & 18.00 &       & 3730 & 6180 & 21.9 \\
Q1554+3749     &  2.664 & 18.19 &       & 3240 & 4770 & 13.2 \\
HS1700+6416    &  2.722 & 16.13 &       & 3730 & 6180 & 66.2 \\
Q1759+7539     &  3.050 & 16.50 &       & 3580 & 5050 & 30.9 \\
Q1937-1009     &  3.806 &       & 16.7  & 3890 & 7450 & 76.9 \\
HS1946+7658    &  3.051 & 16.20 &       & 3890 & 6300 & 136  \\
Q2223+2024     &  3.560 &       & 18.5  & 4120 & 6520 & 12.9 \\
Q2344+1228     &  2.763 & 17.50 &       & 3410 & 4940 &  8.1 \\
\enddata
\tablenotetext{a}{Quasar names are based on B1950 coordinates.}
\tablenotetext{b}{$V$ magnitude from V\'eron-Cetty \& V\'eron (2003).}
\tablenotetext{c}{$R$ magnitude from USNO-A2.0 Catalog (Monet et
  al. 1998), except for Q1330$+$0108, whose R magnitude comes from the
  USNO-B Catalog (Monet et al. 2003).}
\tablenotetext{d}{Lower limit of the observed quasar spectrum.}
\tablenotetext{e}{Upper limit of the observed quasar spectrum.}
\tablenotetext{f}{S/N ratio at the center of each spectrum.}
\tablenotetext{g}{This lensed quasar is amplified by a factor of
  $\sim$3.1 (Barvainis \& Ivison 2002).}
\tablenotetext{h}{This lensed quasar is amplified by a factor of
  15.38 (Kormann et al. 1994).}

\end{deluxetable}


\begin{deluxetable}{lcccccccccclc}
\rotate
\tabletypesize{\scriptsize}
\setlength{\tabcolsep}{0.02in}
\tablecaption{Properties of 86 \hi\ systems\label{tab:2}}
\tablewidth{0pt}
\tablehead{
\colhead{(1)} &
\colhead{(2)} &
\colhead{(3)} &
\colhead{(4)} &
\colhead{(5)} &
\colhead{(6)} &
\colhead{(7)} &
\colhead{(8)} &
\colhead{(9)} &
\colhead{(10)} &
\colhead{(11)} &
\colhead{(12)} &
\colhead{(13)} \\
\colhead{quasar} &
\colhead{$z_{abs}$} &
\colhead{$\log N_{1}$ $^a$} &
\colhead{$\log N_{2}$ $^b$} &
\colhead{$N_{1}/N_{2}$} &
\colhead{S/N(Ly-1)} &
\colhead{S/N(Ly-2)} &
\colhead{S/N(Ly-5)} &
\colhead{S/N(Ly-10)} &
\colhead{$n_{1000}$ $^c$} &
\colhead{$n_{sys}$ $^d$} &
\colhead{status $^e$} &
\colhead{reference$^f$} \\
\colhead{} &
\colhead{} &
\colhead{(cm$^{-2}$)} &
\colhead{(cm$^{-2}$)} &
\colhead{} &
\colhead{} &
\colhead{} &
\colhead{} &
\colhead{} &
\colhead{} &
\colhead{} &
\colhead{} &
\colhead{} \\
}
\startdata
Q0004+1711       & 2.8284 & 15.51 & 14.46 & 0.0896 &  18 & 9.6  & 2.6 & 1.8 & 10  & 2 & $V_{5000}$       &       \\
                 & 2.8540 & 15.75 & 14.94 & 0.1546 &  27 & 8.5  & 3.7 & 2.2 & 17  &...& A, $V_{5000}$    &       \\
                 & 2.8707 & 19.93 & 16.03 & 0.0001 &  27 &  11  & 3.4 & 2.0 & 9   &...& A, C, $V_{5000}$ & 1     \\
Q0014+8118       & 2.7989 & 18.30 & 18.02 & 0.5282 &  63 & 5.2  & ... & ... & 11  & 3 &                  & 1     \\
                 & 2.9090 & 16.09 & 15.60 & 0.3221 &  88 &  11  & 2.0 & ... & 18  & 4 &                  & 1     \\
                 & 3.2277 & 15.33 & 15.28 & 0.8855 &  93 &  41  & 9.3 & 4.2 & 18  & 7 &                  & 1     \\
                 & 3.3212 & 16.60 & 16.24 & 0.4438 & 154 &  48  & 14  & 9.0 & 16  & 8 & $V_{5000}$       & 1     \\
Q0054-2824       & 3.2370 & 15.56 & 15.18 & 0.4108 &  17 & 8.0  & ... & ... & 14  & 7 &                  &       \\
                 & 3.3123 & 16.64 & 14.91 & 0.0184 &  25 & 4.9  & ... & ... & 16  & 4 &                  &       \\
                 & 3.4488 & 16.67 & 15.21 & 0.0346 &  32 &  12  & 3.1 & 2.2 & 17  & 9 &                  &       \\
                 & 3.5113 & 15.89 & 14.84 & 0.0899 &  50 &  14  & 4.9 & 3.2 & 17  & 5 &                  & 1     \\
                 & 3.5805 & 17.44 & 15.94 & 0.0318 &  96 &  16  & 6.3 & 5.9 & 21  & 6 & $V_{5000}$       & 1     \\
Q0119+1432       & 2.4299 & 15.93 & 15.14 & 0.1646 &  39 &  11  & ... & ... & 6   & 4 &                  &       \\
                 & 2.5688 & 16.39 & 14.62 & 0.0169 &  41 &  16  & 4.4 & ... & 11  & 5 &                  &       \\
                 & 2.6632 & 19.37 & 15.82 & 0.0003 &  49 &  25  & 7.6 & 5.7 & 11  &...& C                &       \\
HE0130-4021      & 2.8581 & 15.15 & 14.84 & 0.4902 &  59 & 7.2  & ... & ... & 19  & 4 &                  &       \\
Q0241-0146       & ...    &   ... &   ... &    ... &  ...&  ... & ... & ... & ... &...&                  &       \\
Q0249-2212       & 2.6745 & 19.01 & 14.28 & 0.0000 &  15 &  4.8 & ... & ... & 9   &...& C                & 1     \\
                 & 2.9401 & 17.23 & 14.65 & 0.0026 &  11 &  7.8 & 2.4 & 3.4 & 17  & 4 &                  & 1     \\
HE0322-3213      & 3.0812 & 15.68 & 14.86 & 0.1515 &  26 &   15 & 6.2 & ... & 17  & 3 &                  &       \\
                 & 3.1739 & 19.43 & 14.18 & 0.0000 &  27 &   18 &  11 & 7.5 & 9   &...& A, C             &       \\
                 & 3.1960 & 16.61 & 15.74 & 0.1345 &  36 &   22 & 8.7 & 6.5 & 15  &...& A                &       \\
                 & 3.3169 & 16.16 & 15.33 & 0.1475 & 103 &   25 &  13 &  12 & 11  &...&$V_{1000}$        &       \\
Q0336-0143       & ...    &   ... &   ... &    ... &  ...&  ... & ... & ... & ... &...&                  &       \\
Q0450-1310       & ...    &   ... &   ... &    ... &  ...&  ... & ... & ... & ... &...&                  &       \\
Q0636+6801       & 2.6825 & 15.57 & 15.02 & 0.2837 &  64 &   19 & ... & ... & 15  & 3 &                  & 1     \\
                 & 2.8685 & 15.85 & 14.49 & 0.0431 &  87 &   41 & 7.3 & ... & 10  & 3 &                  & 1     \\
                 & 2.9039 & 18.22 & 15.45 & 0.0017 &  60 &   42 & 13  & 7.8 & 19  & 6 &                  & 1     \\
                 & 3.0135 & 15.79 & 14.95 & 0.1465 & 105 &   27 & 17  &  16 & 19  & 3 & D                & 1     \\
                 & 3.0675 & 15.28 & 14.30 & 0.1054 & 117 &   44 & 23  &  13 & 18  & 6 &                  & 1     \\
Q0642+4454       & 2.9726 & 17.36 & 14.68 & 0.0021 &  22 & 7.2  & ... & ... & 18  & 3 & D                & 1     \\
                 & 3.1230 & 19.48 & 17.54 & 0.0116 &  23 & 8.9  & ... & ... & 11  &...& C                & 1     \\
                 & 3.1922 & 15.27 & 14.48 & 0.1640 &  28 & 15   & 1.3 & ... & 18  & 3 &                  & 1     \\
                 & 3.2290 & 15.52 & 15.37 & 0.7158 &  27 & 15   & 3.4 & ... & 19  &...& A                &       \\
                 & 3.2476 & 16.55 & 15.37 & 0.0669 &  29 & 16   & 4.0 & ... & 18  &...& A, B             & 1     \\
                 & 3.3427 & 15.40 & 14.84 & 0.2744 &  20 & 17   & 7.3 & 4.3 & 15  &...& B, $V_{5000}$    & 1     \\
HS0757+5218      & 2.7261 & 15.46 & 14.98 & 0.3360 &  35 & 11   & ... & ... & 10  & 2 &                  &       \\
                 & 2.8922 & 18.34 & 14.90 & 0.0004 &  25 & 18   & 1.4 & ... & 13  & 1 &                  &       \\
                 & 3.0398 & 19.82 & 16.74 & 0.0008 &  30 & 25   &  11 & 6.7 & 10  &...& C                &       \\
Q0805+0441       & 2.7719 & 15.30 & 15.14 & 0.6906 &  29 & 5.8  & ... & ... & 25  & 7 &                  & 1     \\
                 & 2.8113 & 15.99 & 14.88 & 0.0765 &  37 & 8.3  & ... & ... & 17  & 4 &                  & 1     \\
Q0831+1248       & 2.7300 & 15.74 & 14.07 & 0.0212 &  57 & 13   & ... & ... & 11  &...& $V_{1000}$       &       \\
HE0940-1050      & 2.8283 & 16.41 & 16.05 & 0.4305 &  52 & 12   & ... & ... & 20  & 15&                  &       \\
                 & 2.8610 & 17.06 & 14.53 & 0.0029 &  56 & 18   & 2.6 & ... & 18  & 4 &                  &       \\
                 & 2.9174 & 15.92 & 15.35 & 0.2669 &  63 & 21   & 6.2 & ... & 19  & 5 &                  &       \\
                 & 3.0387 & 15.55 & 13.84 & 0.0196 &  91 & 27   & 7.6 & 6.7 & 10  & 3 & $V_{5000}$       &       \\
Q1009+2956       & 2.1432 & 17.82 & 15.33 & 0.0032 &  71 & 17   & ... & ... & 9   & 6 &                  &       \\
                 & 2.4069 & 18.80 & 14.25 & 0.0000 & 126 & 36   & 9.6 & ... & 9   &...& A                & 1     \\
                 & 2.4292 & 17.34 & 14.53 & 0.0015 & 109 & 45   & 14  & 1.7 & 18  &...& A                &       \\
                 & 2.5037 & 17.26 & 15.49 & 0.0167 & 131 & 53   & 21  & 15  & 14  & 4 &                  & 1     \\
Q1017+1055       & 2.9403 & 15.56 & 14.49 & 0.0844 &  12 & 8.2  & ... & ... & 11  & 2 &                  & 1     \\
                 & 3.0096 & 15.98 & 14.80 & 0.0658 &  35 & 6.9  & ... & ... & 21  & 5 &                  & 1     \\
                 & 3.0548 & 17.06 & 15.54 & 0.0302 &  25 & 8.9  & ... & ... & 18  & 11&                  & 1     \\
                 & 3.1120 & 15.26 & 15.01 & 0.5536 &  43 & 11   & ... & ... & 26  & 7 & $V_{5000}$       & 1     \\
Q1055+4611       & 3.8252 & 15.98 & 15.64 & 0.4603 &  53 & 37   & 15  & ... & 26  &...& A                &       \\
                 & 3.8495 & 16.74 & 16.04 & 0.1997 &  31 & 35   & 14  & 4.4 & 23  &...& A                &       \\
                 & 3.9343 & 17.30 & 16.32 & 0.1035 &  40 & 34   & 22  & 13  & 25  &...& B                &       \\
HS1103+6416      & ...    &   ... &   ... &    ... &  ...& ...  & ... & ... & ... &...&                  &       \\
Q1107+4847       & 2.7243 & 16.58 & 13.92 & 0.0022 &  38 & 7.5  & ... & ... & 12  & 3 & D                &       \\
                 & 2.7629 & 19.13 & 17.51 & 0.0239 &  43 & 12   & ... & ... & 12  &...& C                & 1     \\
                 & 2.8703 & 15.25 & 14.76 & 0.3226 &  83 & 18   & ... & ... & 17  & 7 &                  &       \\
Q1157+3143       & 2.7710 & 17.63 & 14.56 & 0.0009 &  69 & 22   & ... & ... & 13  & 3 &                  &       \\
                 & 2.8757 & 15.66 & 15.54 & 0.7713 &  85 & 28   & ... & ... & 19  & 9 &                  &       \\
                 & 2.9437 & 17.44 & 17.16 & 0.5282 &  99 & 40   & ... & ... & 18  & 5 & $V_{5000}$       &       \\
Q1208+1011       & 3.3846 & 17.35 & 15.04 & 0.0049 &  24 & 18   & 6.1 & 3.0 & 19  & 6 &                  &       \\
                 & 3.4596 & 16.88 & 16.03 & 0.1430 &  22 & 18   & 8.4 & 5.5 & 19  & 10&                  &       \\
                 & 3.5195 & 16.15 & 15.73 & 0.3802 &  26 & 19   & 11  & 7.7 & 24  & 6 &                  &       \\
                 & 3.7206 & 15.48 & 14.65 & 0.1485 &  27 & 19   & 14  & 12  & 21  & 6 &                  &       \\
Q1244+1129       & 2.9318 & 15.97 & 14.87 & 0.0784 &  31 & 14   & 5.0 & 3.5 & 17  & 3 & $V_{5000}$       &       \\
Q1251+3644       & 2.8684 & 15.82 & 14.03 & 0.0161 &  63 & 20   & ... & ... & 14  & 3 &                  &       \\
Q1330+0108       & ...    &   ... &   ... &    ... &  ...&  ... & ... & ... & ... &...&                  &       \\
Q1334-0033       & 2.7572 & 15.40 & 14.24 & 0.0693 &  60 & 9.0  & ... & ... & 15  & 3 & $V_{5000}$       &       \\
Q1337+2832       & 2.4336 & 18.92 & 16.32 & 0.0025 &  60 & 14   & 2.5 & ... & 14  & 8 &                  &       \\
                 & 2.5228 & 15.81 & 14.44 & 0.0423 & 161 & 22   & 6.9 & 2.1 & 18  & 5 & $V_{5000}$       &       \\
Q1422+2309       & 3.3825 & 16.53 & 16.33 & 0.6427 & 389 & 278  & 83  & 54  & 19  & 4 &                  &       \\
                 & 3.5362 & 15.94 & 15.83 & 0.7691 & 462 & 370  &170  & 70  & 22  &...& B, $V_{5000}$    &       \\
Q1425+6039       & 2.7700 & 19.37 & 16.20 & 0.0007 &  38 & 6.3  & ... & ... & 17  &...& C, D             &       \\
                 & 2.8258 & 20.00 & 19.68 & 0.4716 &  33 & 9.0  & ... & ... & 7   &...& C                & 1     \\
                 & 3.0671 & 16.20 & 14.90 & 0.0496 &  94 & 13   & 4.4 & 1.1 & 16  & 3 &                  &       \\
                 & 3.1356 & 16.66 & 16.15 & 0.3107 & 184 &127   & 6.4 & 2.7 & 17  &...& B, $V_{5000}$    &       \\
Q1442+2931       & ...    &  ...  &  ...  &    ... &  ...& ...  & ... & ... & ... &...&                  &       \\
Q1526+6701       & 2.9751 & 15.12 & 15.11 & 0.9808 &  24 & 7.5  & 4.5 & 3.0 & 13  & 5 & $V_{5000}$       &       \\
Q1548+0917       & ...    &  ...  &  ...  &    ... &  ...& ...  & ... & ... & ... &...&                  &       \\
Q1554+3749       & 2.6127 & 17.97 & 14.45 & 0.0003 &  43 & 7.5  & 2.7 & 1.5 & 11  & 3 & $V_{5000}$       &       \\
HS1700+6416      & ...    &  ...  &  ...  &    ... &  ...& ...  & ... & ... & ... &...&                  &       \\
Q1759+7529       & 2.7953 & 15.26 & 14.92 & 0.4584 &  44 & 22   & ... & ... & 16  & 5 &                  & 1     \\
                 & 2.8493 & 17.44 & 15.60 & 0.0145 &  49 & 27   & 7.9 & ... & 16  & 6 &                  & 1     \\
                 & 2.9105 & 19.90 & 17.62 & 0.0052 &  60 & 32   & 10  & 6.8 & 15  &...& C                & 1     \\
Q1937-1009       & 3.5725 & 17.94 & 15.89 & 0.0089 & 147 & 61   & 27  & 20  & 20  & 4 &                  & 1     \\
HS1946+7658      & 3.0498 & 17.45 & 14.42 & 0.0009 & 266 & 44   & ... & ... & 10  &...& D, $V_{1000}$    & 1     \\
Q2223+2024       & ...    &  ...  &  ...  &    ... &  ...& ...  & ... & ... & ... &...&                  &       \\
Q2344+2024       & 2.4261 & 18.46 & 15.18 & 0.0005 &  16 & 6.8  & ... & ... & 8   & 3 &                  & 1     \\
                 & 2.6356 & 15.45 & 14.03 & 0.0382 &  21 &  11  & 1.5 & ... & 7   & 2 &                  & 1     \\
                 & 2.7107 & 16.64 & 15.68 & 0.1092 &  31 &  13  & 4.8 & 1.7 & 12  & 6 & $V_{5000}$       & 1     \\
                 & 2.7469 & 16.67 & 16.27 & 0.4039 &  34 &  14  & 6.7 & 3.1 & 19  & 10& $V_{5000}$       &       \\
\enddata
\tablenotetext{a}{\hi\ The largest column density.}
\tablenotetext{b}{\hi\ The second largest column density within $\pm
  1000$ ~\kms\ of the largest column density.}
\tablenotetext{c}{Number of \hi\ lines within $\pm$1000~\kms\ of the
  main component.}
\tablenotetext{d}{Number of HDLs (see \S\ 3.3).}
\tablenotetext{e}{A : centers of \hi\ systems are separated  
                       by less that 1000~\kms,
                  B : there are gaps in the spectrum within 1000~\kms\ of the
                      main component,
                  C : normalization of spectrum is not good because of
                      strong absorption with \lognhi\ $>$ 19,
                  D : candidate for a quasar intrinsic system (Misawa
                      et al. 2007),
                  $V_{1000}$ : velocity difference from the quasar
                      emission redshift is smaller than 1000~\kms,
                  $V_{5000}$ : velocity difference from the quasar
                      emission redshift is smaller than 5000~\kms.}
\tablenotetext{f}{1 : Listed in NED or literature in 2002 (P\'eroux et
                      al. 2001; Storrie-Lombardi et al. 1996;
                      Petitjean, Rauch, \& Carswell 1994; Lu et
                      al. 1993; Steidel \& Sargent 1992; Lanzetta et
                      al. 1991; Barthel, Tytler, \& Thomson 1990;
                      Steidel 1990a,b; SBS; SSB).}
\end{deluxetable}


\begin{deluxetable}{cccc}
\tabletypesize{\scriptsize}
\tablecaption{Sub-samples of H~I lines for statistical analysis\label{tab:3}}
\tablewidth{0pt}
\tablehead{
\colhead{(1)} &
\colhead{(2)} &
\colhead{(3)} &
\colhead{(4)} \\
\colhead{Sub-sample$^{a}$} &
\colhead{$N_{sys}$ $^{b}$} &
\colhead{$N_{line}$ $^{c}$} &
\colhead{Criteria} \\
}
\startdata
S0          &  86 & 1339 & All \hi\ systems and lines \\
\hline
S1          &  61 &  973 & \hi\ systems meeting the conditions in \S\ 2 $^{d}$ \\
\hline
S2a         &  48 &  767 & $\Delta v$(\zem $-$ \zabs) $>$    5000 \kms \\
S2b         &  13 &  206 & $\Delta v$(\zem $-$ \zabs) $\leq$ 5000 \kms \\
\hline
S3a         &  34 &  554 & S/N ratio at \lya\ is larger than 40 \\
S3b         &  17 &  280 & S/N ratio at \lya\ is larger than 70 \\
\hline
S4a         &  30 &  419 & \zabs\ $<$ 2.9 \\
S4b         &  31 &  554 & \zabs\ $\geq$ 2.9 \\
\hline
S5$_{1213}$ & ... &  244 & $12 < \log N_{HI} < 13$ \\
S5$_{1214}$ & ... &  716 & $12 < \log N_{HI} < 14$ \\
S5$_{1215}$ & ... &  866 & $12 < \log N_{HI} < 15$ \\
S5$_{1216}$ & ... &  885 & $12 < \log N_{HI} < 16$ \\
S5$_{1319}$ & ... &  728 & $13 < \log N_{HI} < 19$ \\
S5$_{1419}$ & ... &  256 & $14 < \log N_{HI} < 19$ \\
S5$_{1519}$ & ... &  106 & $15 < \log N_{HI} < 19$ \\
S5$_{1619}$ & ... &   58 & $16 < \log N_{HI} < 19$ \\
\enddata
\tablenotetext{a}{Sub-samples S2a -- S5$_{1619}$ are all derived from
  a sample S1.}
\tablenotetext{b}{Number of \hi\ systems, i.e. lines with \lognhi\ $>$
  15 \cmm .}
\tablenotetext{c}{Number of \hi\ lines, i.e. all H~I lines within $\pm
  1000$~\kms\ of the systems.}
\tablenotetext{d}{Should satisfy four conditions; (i) separated from
  the quasar by more than 1000~\kms, (ii) column density of the main
  component is \lognhi\ $<$ 19 \cmm , (iii) there are no gaps in the
  HIRES spectrum near \lya , and (iv) there are no other \hi\ systems
  within $\pm$2000~\kms.}
\end{deluxetable}


\begin{deluxetable}{cccc}
\tabletypesize{\scriptsize}
\tablecaption{Parameters of the column density distribution\label{tab:4}}
\tablewidth{0pt}
\tablehead{
\colhead{(1)} &
\colhead{(2)} &
\colhead{(3)} &
\colhead{(4)} \\
\colhead{sub-sample} &
\colhead{$\beta$ $^{a}$} &
\colhead{$A$ $^{b}$} &
\colhead{$\Delta z$ $^{c}$} \\
}
\startdata
S1 (all lines)                  & 1.398 $\pm$ 0.025 & 7.389 $\pm$ 0.407 & 1.606 \\
\hline
S2a ($\Delta v > 5000$ km/s)    & 1.390 $\pm$ 0.027 & 7.262 $\pm$ 0.431 & 1.266 \\
S2b ($\Delta v \leq 5000$ km/s) & 1.360 $\pm$ 0.048 & 6.801 $\pm$ 0.753 & 0.341 \\
\hline
S3a (S/N $>$ 40)                & 1.368 $\pm$ 0.034 & 6.937 $\pm$ 0.552 & 0.887 \\
S3b (S/N $>$ 70)                & 1.350 $\pm$ 0.040 & 6.660 $\pm$ 0.615 & 0.453 \\
\hline
S4a ($z$ $<$ 2.9)               & 1.343 $\pm$ 0.031 & 6.537 $\pm$ 0.494 & 0.741 \\
S4b ($z$ $\geq$ 2.9)            & 1.439 $\pm$ 0.031 & 8.015 $\pm$ 0.497 & 0.865 \\
\hline
Kirkman \& Tytler (1997)        & 1.5               & 8.79              &       \\
Petitjean et al. (1993)         & 1.46              & 8.08              &       \\
\enddata
\tablenotetext{a}{Best fit value and 1$\sigma$ error of $\beta$ in
  eqn.(\ref{eqn:1}).}
\tablenotetext{b}{Best fit value and 1$\sigma$ error of $A$ in
  eqn.(\ref{eqn:1}).}
\tablenotetext{c}{Total redshift width of sub-sample.}
\end{deluxetable}


\begin{deluxetable}{cccc}
\tabletypesize{\scriptsize}
\tablecaption{ K-S test for column density\label{tab:5}}
\tablewidth{0pt}
\tablehead{
\colhead{(1)} &
\colhead{(2)} &
\colhead{(3)} \\
\colhead{sub-samples} &
\colhead{$D$} &
\colhead{Prob$^{a}$} \\
\colhead{} &
\colhead{} &
\colhead{(\%)} \\
}
\startdata
S2a / S2b & 0.063 & 52.6 \\
S3a / S3b & 0.073 & 21.1 \\
S4a / S4b & 0.055 & 44.5 \\
\enddata
\tablenotetext{a}{Probability that the two distributions were drawn
  from the same parent population.}
\end{deluxetable}


\begin{deluxetable}{cccc}
\tabletypesize{\scriptsize}
\tablecaption{K-S test for Doppler parameter\label{tab:6}}
\tablewidth{0pt}
\tablehead{
\colhead{(1)} &
\colhead{(2)} &
\colhead{(3)} \\
\colhead{sub-samples} &
\colhead{$D$} &
\colhead{Prob$^{a}$} \\
\colhead{} &
\colhead{} &
\colhead{(\%)} \\
}
\startdata
S2a / S2b & 0.116 &  2.4 \\
S3a / S3b & 0.036 & 97.1 \\
S4a / S4b & 0.063 & 28.9 \\
\enddata
\tablenotetext{a}{Probability that the two distributions were drawn
  from the same parent population.}
\end{deluxetable}


\begin{deluxetable}{ccccc}
\tabletypesize{\scriptsize}
\tablecaption{Sub-samples separated into HDLs and LDLs\label{tab:7}}
\tablewidth{0pt}
\tablehead{
\multicolumn{2}{c}{(1)} &
\multicolumn{1}{c}{(2)} &
\multicolumn{1}{c}{(3)} &
\multicolumn{1}{c}{(4)} \\
\multicolumn{2}{c}{Sub-sample} &
\multicolumn{1}{c}{$N_{sys}$ $^{a}$} &
\multicolumn{1}{c}{$N_{line}$ $^{b}$} &
\multicolumn{1}{c}{Criteria}
}
\startdata
S1  & HDLs &  61 &  306 & \hi\ systems meeting the conditions in \S\ 4.1 \\
    & LDLs &  61 &  667 & \hi\ systems meeting the conditions in \S\ 4.1 \\
\hline
S2a & HDLs &  48 &  240 & $\Delta v$(\zem $-$ \zabs) $>$    5000 \kms \\
    & LDLs &  48 &  527 & $\Delta v$(\zem $-$ \zabs) $>$    5000 \kms \\
S2b & HDLs &  13 &   66 & $\Delta v$(\zem $-$ \zabs) $\leq$ 5000 \kms \\
    & LDLs &  13 &  140 & $\Delta v$(\zem $-$ \zabs) $\leq$ 5000 \kms \\
\hline
S4a & HDLs &  30 &  143 & \zabs\ $<$ 2.9 \\
    & LDLs &  30 &  276 & \zabs\ $<$ 2.9 \\
S4b & HDLs &  31 &  163 & \zabs\ $\geq$ 2.9 \\
    & LDLs &  31 &  391 & \zabs\ $\geq$ 2.9 \\
\enddata
\tablenotetext{a}{Number of \hi\ systems}
\tablenotetext{b}{Number of \hi\ lines}
\end{deluxetable}


\begin{deluxetable}{ccccccccccc}
\tabletypesize{\scriptsize}
\tablecaption{Parameters of column density distribution\label{tab:8}}
\tablewidth{0pt}
\tablehead{
\multicolumn{2}{c}{(1)} &
\multicolumn{1}{c}{(2)} &
\multicolumn{1}{c}{(3)} &
\multicolumn{1}{c}{(4)} \\
\multicolumn{2}{c}{sub-sample} &
\multicolumn{1}{c}{$\beta$ $^{a}$} &
\multicolumn{1}{c}{$A$ $^{b}$} &
\multicolumn{1}{c}{$\Delta z$ $^{c}$} \\
}
\startdata
S1                          & HDLs            & 1.269 $\pm$ 0.034 & 5.736 $\pm$ 0.545 & 0.415 \\
(all lines)                 & LDLs            & 1.589 $\pm$ 0.075 & 10.04 $\pm$ 1.045 & 1.192 \\
\hline
S2a                         & HDLs            & 1.264 $\pm$ 0.032 & 5.663 $\pm$ 0.512 & 0.332 \\
($\Delta v > 5000$ km/s)    & LDLs            & 1.526 $\pm$ 0.054 & 9.172 $\pm$ 0.757 & 0.933 \\
\hline
S2b                         & HDLs            & 1.167 $\pm$ 0.047 & 4.203 $\pm$ 0.739 & 0.082 \\
($\Delta v \leq 5000$ km/s) & LDLs            & 1.897 $\pm$ 0.163 & 14.25 $\pm$ 2.278 & 0.259 \\
\hline
S4a                         & HDLs            & 1.223 $\pm$ 0.031 & 5.050 $\pm$ 0.496 & 0.190 \\
($z$ $<$ 2.9)               & LDLs            & 1.712 $\pm$ 0.055 & 11.64 $\pm$ 0.766 & 0.551 \\
\hline
S4b                         & HDLs            & 1.283 $\pm$ 0.044 & 5.933 $\pm$ 0.693 & 0.227 \\
($z$ $\geq$ 2.9)            & LDLs            & 1.517 $\pm$ 0.091 & 9.102 $\pm$ 1.274 & 0.638 \\
\hline
Petitjean et al. (1993)     & (HDLs $+$ LDLs) & 1.46              & 8.08              &       \\
Kirkman \& Tytler (1997)    & (LDLs)          & 1.5               & 8.79              &       \\
\enddata
\tablenotetext{a}{Best fit value and 1$\sigma$ error of $\beta$ in
  eqn.(\ref{eqn:1}).}
\tablenotetext{b}{Best fit value and 1$\sigma$ error of $A$ in
  eqn.(\ref{eqn:1}).}
\tablenotetext{c}{Total redshift width of sub-sample.}
\end{deluxetable}


\begin{deluxetable}{ccc}
\tabletypesize{\scriptsize}
\tablecaption{K-S test for Doppler parameter\label{tab:9}}
\tablewidth{0pt}
\tablehead{
\colhead{(1)} &
\colhead{(2)} &
\colhead{(3)} \\
\colhead{sub-samples} &
\colhead{$D$} &
\colhead{Prob$^{a}$} \\
\colhead{} &
\colhead{} &
\colhead{(\%)} \\
}
\startdata
HDLs (S1)           / LDLs (S1)                                        & 0.049 & 67.9 \\
\hline
HDLs ($\Delta v > 5000$ \kms) / LDLs ($\Delta v > 5000$ \kms)          & 0.084 & 18.8 \\
HDLs ($\Delta v \leq 5000$ \kms) / LDLs ($\Delta v \leq 5000$ \kms)    & 0.158 & 19.4 \\
\hline
HDLs ($\Delta v > 5000$ \kms) / HDLs ($\Delta v \leq 5000$ \kms)       & 0.091 & 76.6 \\
LDLs ($\Delta v > 5000$ \kms) / LDLs ($\Delta v \leq 5000$ \kms)       & 0.143 &  2.0 \\
\hline
HDLs ($z < 2.9$)    / LDLs ($z < 2.9$)                                 & 0.095 & 34.3 \\
HDLs ($z \geq 2.9$) / LDLs ($z \geq 2.9$)                              & 0.086 & 34.6 \\
\hline
HDLs ($z < 2.9$)    / HDLs ($z \geq 2.9$)                              & 0.134 & 11.7 \\
LDLs ($z < 2.9$)    / LDLs ($z \geq 2.9$)                              & 0.041 & 94.7 \\
\enddata
\tablenotetext{a}{Probability that the two distributions were drawn
  from the same parent population.}
\end{deluxetable}


\clearpage


\begin{figure}
\centerline{
\includegraphics[width=12cm]{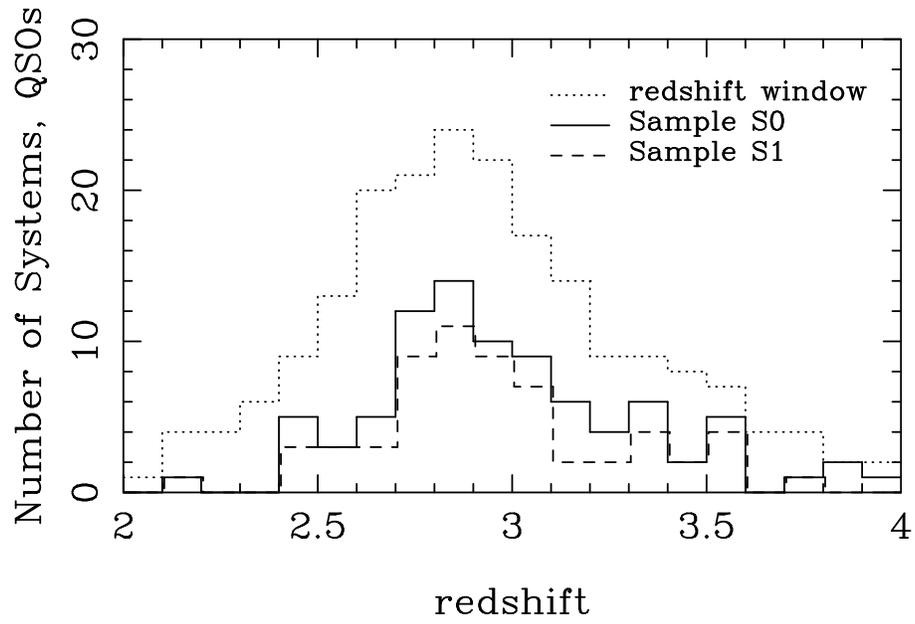}
}
\caption{The solid and dashed histograms represent the numbers of \hi\
 systems in samples S0 and S1 as a function of redshift. The
 dot-dashed histogram is the number of quasars in which absorption
 lines could have been detected (namely the number of spectrum
 windows).\label{fig:1}}
\end{figure}

\clearpage


\begin{figure}
\centerline{
\includegraphics[width=12cm]{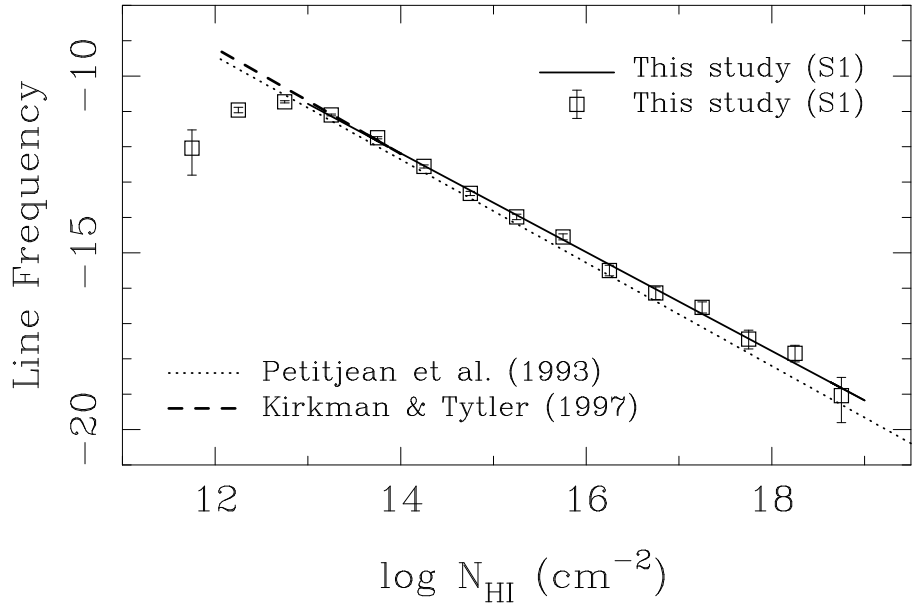}
}
\caption{Observed column density distribution per unit redshift and
 unit column density for sample S1. The \hi\ lines have been binned
 into intervals of 0.5 in \lognhi. The open squares and vertical bars
 represent the observed data and 1$\sigma$ errors. The solid line is
 the best fit power law for our study. Dashed and dotted lines are the
 best fit power laws in the range of 12 $<$ \lognhi\ $<$ 14 (KT97) and
 12 $<$ \lognhi\ $<$ 22 (Petitjean et al. 1993).\label{fig:2}}
\end{figure}



\begin{figure}
\centerline{
\includegraphics[width=8cm]{f3a.eps}
\hspace{0.5cm}
\includegraphics[width=8cm]{f3b.eps}
}
\caption{Observed Doppler parameter distributions for sub-samples S1
 and S$_{1213}$. The \hi\ lines have been binned into intervals of
 $\Delta b$ = 5~\kms. The open squares and vertical bars represent the
 observed data with their 1$\sigma$ errors. Dashed and dotted lines
 are the input data with truncated Gaussian distribution profiles for
 H95 ($b_{0}$ = 28~\kms, $\sigma_{b}$ = 10~\kms, and $b_{cut}$ =
 20~\kms) and L96 ($b_{0}$ = 23~\kms, $\sigma_{b}$ = 8~\kms, and
 $b_{cut}$ = 15~\kms).\label{fig:3}}
\end{figure}

\clearpage


\begin{figure}
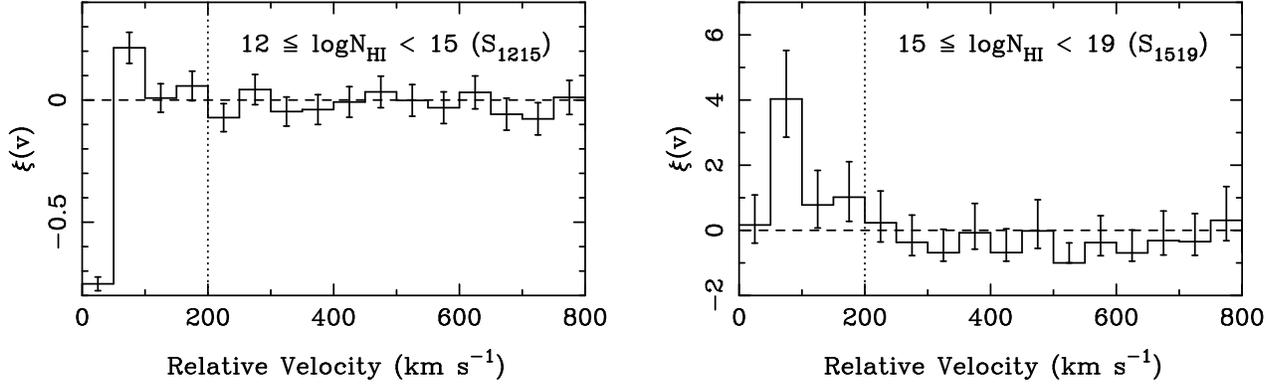

\centerline{
\includegraphics[width=8cm]{f4a.eps}
\hspace{0.5cm}
\includegraphics[width=8cm]{f4b.eps}
}
\caption{Two point correlation functions for sub-samples $S5_{1215}$
 and $S5_{1519}$. The bin size is 50~\kms. Solid histogram and
 vertical bars in each bin represent the value of correlation degree,
 $\xi (v)$, and the Poisson error. Dotted vertical line denotes the
 velocity separation at which the lower 1$\sigma$ deviation of $\xi
 (v)$ first goes below $\xi (v)$ = 0 over $v$ $>$
 50~\kms.\label{fig:4}}
\end{figure}



\begin{figure}
\centerline{
\includegraphics[width=8cm]{f5a.eps}
\hspace{0.5cm}
\includegraphics[width=8cm]{f5b.eps}
}
\caption{Observed Doppler parameter distributions of LDLs for
 sub-samples S2a ($\Delta v$ $>$ 5000~\kms) and S2b ($\Delta v$ $\leq$
 5000~\kms). The \hi\ lines have been binned into intervals of $\Delta
 b$ = 5~\kms. Open square and vertical bars represent observed data
 and the 1$\sigma$ errors. Dashed and dotted lines are the input data
 with truncated Gaussian distribution profiles for H95 ($b_{0}$ =
 28~\kms, $\sigma_{b}$ = 10~\kms, and $b_{min}$ = 20~\kms) and L96
 ($b_{0}$ = 23~\kms, $\sigma_{b}$ = 8~\kms, and $b_{min}$ =
 15~\kms).\label{fig:5}}
\end{figure}

\clearpage


\begin{figure}
\centerline{
\includegraphics[width=12cm]{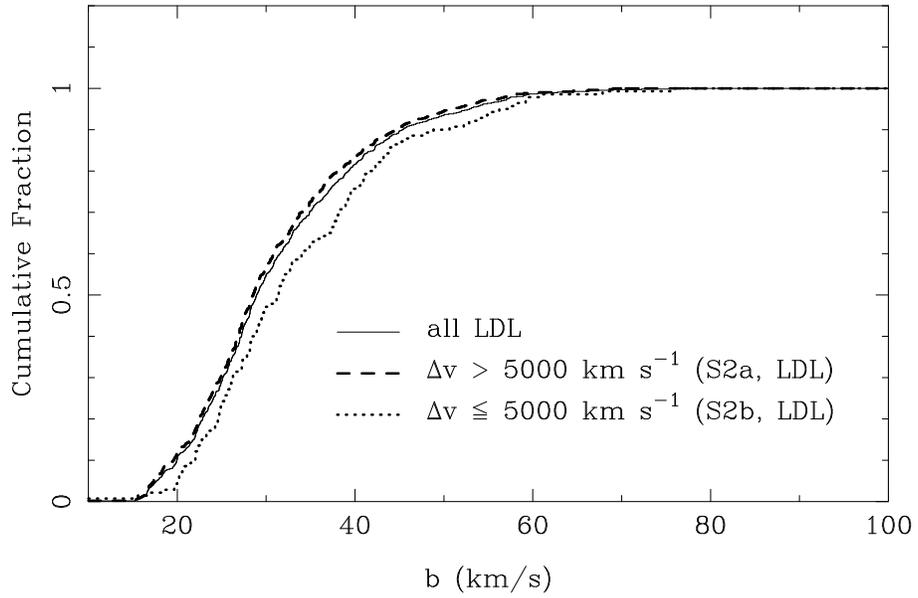}
}
\caption{The cumulative distribution functions for the $b$-values of
LDLs.  The 527 LDLs from $\Delta v$ $>$ 5000~\kms\ (sample S2a) are
shown with the dashed line, the 140 LDLs at $\Delta v$ $\leq$
5000~\kms\ (S2b) with the dotted line, and we show all 667 LDLs (S1)
with the central thin solid line. \lya\ forest lines near to a quasar
tend to have larger Doppler parameters compared with lines far from a
quasar.\label{fig:6}
}
\end{figure}

\clearpage


\begin{figure}
\centerline{
\includegraphics[width=12cm]{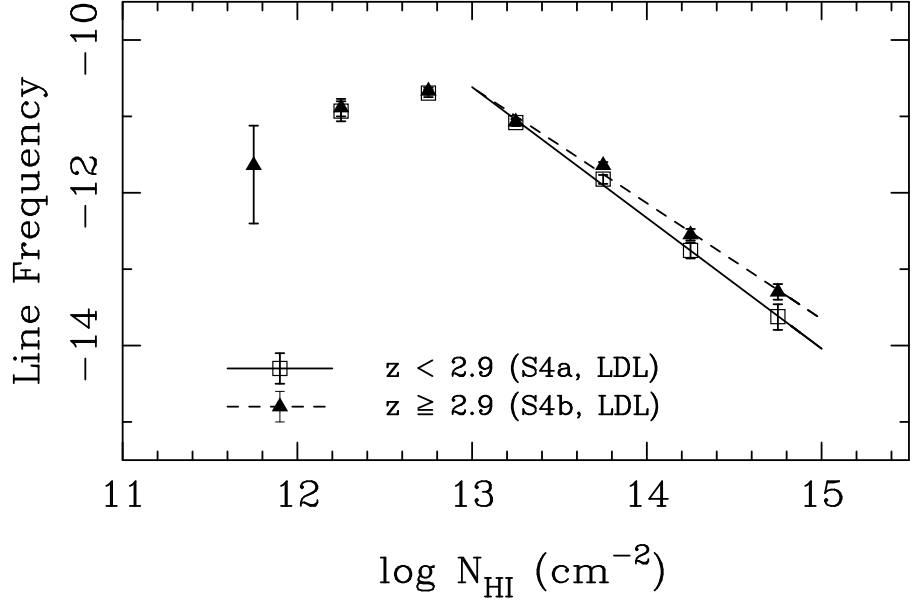}
}
\caption{Column density distributions of LDLs at $z$ $<$ 2.9 (S4a)
 with open squares and solid lines, and at $z$ $\geq$ 2.9 (S4b) with
 filled triangles and dashed lines. The line frequency of stronger
 LDLs (i.e., \lognhi\ $>$ 14.5) at $z$ $<$ 2.9 preferentially
 decreases compared with those of LDLs at $z$ $\geq$ 2.9, while the
 frequency of weaker LDLs does not change with redshift.\label{fig:7}}
\end{figure}



\begin{figure}
\centerline{
\includegraphics[width=12cm]{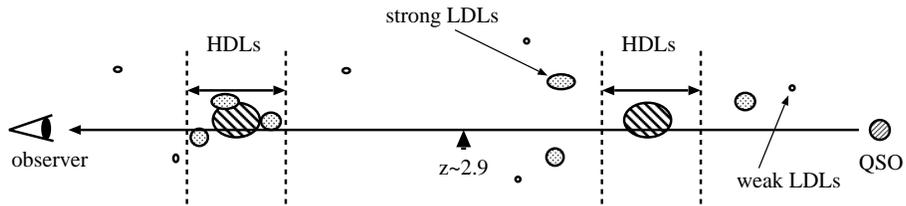}
}
\caption{Cartoon of the distribution of absorbers at $z$ $\geq$ 2.9
 and $z$ $<$ 2.9. Shaded circles are HDL absorbers. Dotted and small
 circles are strong (e.g., \lognhi\ $>$ 14.5) and weak (e.g., \lognhi\
 $<$ 14.5) LDL absorbers, respectively. If strong LDL absorbers would
 gather around HDL absorbers within relative velocity of $\Delta v$
 $<$ 200~\kms, the number of strong LDLs decrease as redshift
 decreases. Such trend is consistent with the concept of the
 hierarchical clustering model.\label{fig:8}}
\end{figure}

\clearpage


\begin{figure}
\centerline{
\includegraphics[width=12cm]{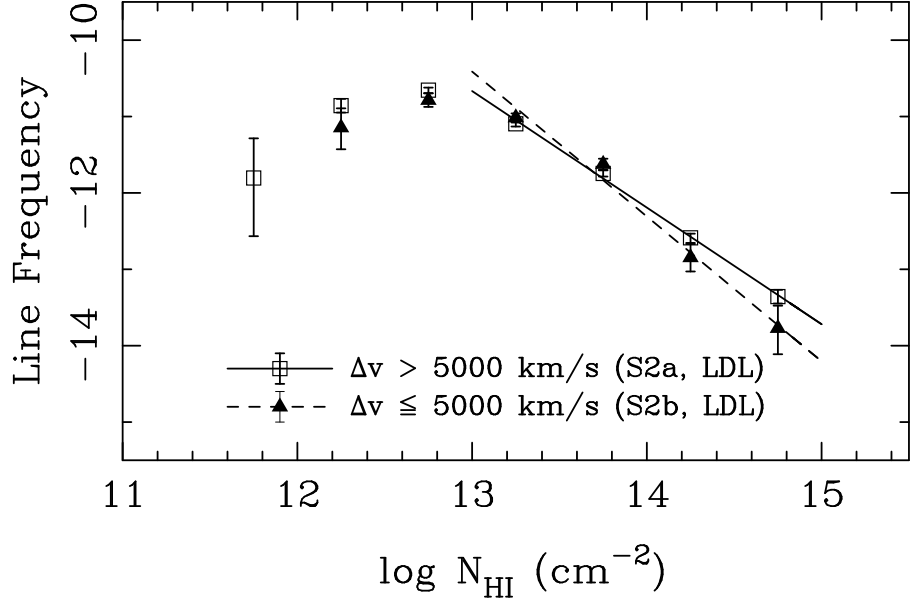}
}
\caption{Column density distributions of LDLs far from the quasars
 with relative radial velocity of $\Delta v$ $>$ 5000~\kms\ (S2a)
 which are shown as open squares and solid lines, and near the quasars
 with $\Delta v$ $\leq$ 5000~\kms\ (S2b) denoted as filled triangles
 and dashed lines. The line frequency of stronger LDLs near the
 quasars preferentially decreases compared with those of LDLs far from
 the quasars, while the frequency of weaker LDLs is not affected by
 the distance from the quasars.\label{fig:9}}
\end{figure}



\begin{figure}
\centerline{
\includegraphics[width=12cm]{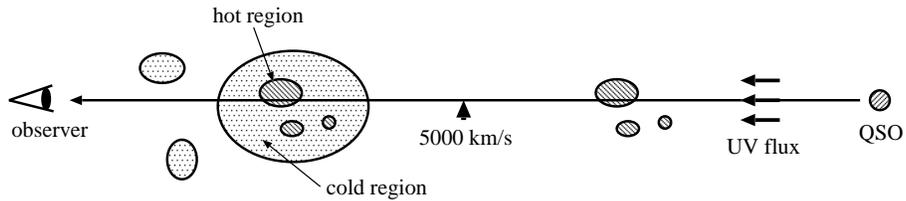}
}
\caption{Cartoon of the distribution of absorbers far from the quasars
 ($\Delta v$ $>$ 5000~\kms) and near the quasars ($\Delta v$ $\leq$
 5000~\kms). Shaded regions are dense regions that are adiabatically
 compressed and heated, while dotted regions are diffuse cold regions.
 If LDL absorbers near the quasars are strongly affected by the UV
 flux from the quasars, \hi\ gas at the diffuse cold regions would be
 preferentially ionized. As a result, stronger LDLs would become to be
 weaker LDLs and only central hot regions are observed, which makes
 the relative frequency of stronger LDLs smaller and the mean $b$
 value of LDLs larger.\label{fig:10}}
\end{figure}



\clearpage

\appendix

\section{Discussion of individual H I systems}

In this section, we describe the results of fitting the 86 \hi\
systems in sample S0. Velocity plots of them with $\pm$ 1000~\kms\
widths for the lowest five orders of Lyman series (i.e., \lya, \lyb,
\lyg, \lyd, and \lye) are presented in Figure~11 as far as they are
accessible.  In Table~10 we give in column (1) ID number; columns (2)
and (3) observed wavelength and velocity shift from the system center;
column (4) absorption redshift; columns (5) and (6) column density
with 1$\sigma$ error; columns (7) and (8) Doppler parameter with
1$\sigma$ error; column (9) line identification. If narrow lines with
$b$ $<$ 15~\kms\ are not identified as specific metal lines, we use
``\ion{M}{1}'' as unidentified lines in the column (10).  Table~10
lists only \ion{H}{1}, \ion{M}{1}, and metal lines that are detected
within $\pm$1000~\kms\ windows of the 81 \hi\ systems. Metal lines in
the \hi\ system windows are neither numerated in the table nor marked
with ticks in Figure~11 because they happen to locate within the \hi\
system windows and they are not physically relate to the \hi\
systems. Important metal absorption lines in the 86 \hi\ systems that
are detected in our spectra are also summarized in a separate table
(Table~11).

\begin{description}

\item{\it Q0004+1711 (\zem\ = 2.890). ---} SSB observed this quasar,
and detected strong \ion{C}{4} and \ion{Si}{4} absorption lines at
\zabs=2.5181 as well as a strong \ion{Mg}{2} line at \zabs\ =
0.8068. We confirm the prominent LLS at \zabs\ = 2.881. We see
\ion{Si}{2}~$\lambda$1260, \ion{Si}{2}~$\lambda$1527,
\ion{C}{2}~$\lambda$1335, and \ion{O}{1}~$\lambda$1302 lines, but no
\ion{C}{4} doublet. Our spectrum has a range of 3510~\AA\ to
5030~\AA. Both \lya\ and \lyb\ are detected at \zabs\ = 2.422 --
2.890.

{\it \zabs\ = 2.8284 ---} Although the spectrum has a range of \lya\
up to Ly13, the S/N ratio is very low (S/N = 18 at \lya, and 1.8 at
Ly10). This system is within 5000~\kms\ of the emission redshift of
the quasar at \zem\ = 2.89.

{\it \zabs\ = 2.8540 ---} This system is also within 5000~\kms\ of the
quasar. Though the spectrum covers the Lyman limit of the system
($\lambda_{limit}$ $\sim$ 3513~\AA), the low S/N ratio of the spectrum
prevented us from measuring this. This system is shifted only
1300~\kms\ blueward of the DLA system at \zabs\ = 2.8707, and \lya\ is
strongly blended with the left wing of the DLA.

{\it \zabs\ = 2.8707 ---} This system was previously detected by
SBS. Most components in the system are blanketed by the wings of the
main component, which has a large column density, \lognhi\ = 19.93,
and rather small Doppler parameter, $b$ = 12.57 \kms. This system is
also within 5000~\kms\ of the quasar.

\item{\it Q0014+8118 (\zem\ = 3.387). ---} This quasar has been well
studied since its discovery in 1983 (Kuhr et al. 1983), as there is a
candidate \ion{D}{1} line at \zabs\ = 3.32. An upper limit on the D/H
ratio was determined to be D/H $<$ 25 -- 60 $\times 10^{-5}$ for this
system (Songaila et al. 1994; Carswell et al. 1994). Rugen \& Hogan
(1996a,b) also detected a \ion{D}{1} line in another LLS at \zabs\ =
2.80 in this quasar. Burles, Kirkman, \& Tytler (1999), however,
claimed that these absorption lines were not primarily due to
\ion{D}{1}, based on their improved spectrum. Our spectrum ranges from
3650~\AA\ to 6080~\AA. Both \lya\ and \lyb\ are detected at \zabs\ =
2.558 -- 3.387.

{\it \zabs\ = 2.7989 ---} The absorption profile around the main
component (the ``central trough'' hereafter) is strongly damped for
\lya, \lyb, and \lyg, which makes it difficult to fit the profile. If
the trough was fit with a single component, the Doppler parameter was
found to be rather large, $b$ $>$ 60~\kms. Fortunately this system has
many \ion{C}{4} and \ion{Si}{4} lines. Therefore the \ion{C}{4} lines
were used as a reference, and the trough was fit with two components
having $b$ = 45 and 33~\kms, respectively. Both components are found
to have high column densities, \lognhi\ $>$ 18. They may be resolved
into narrower components.

{\it \zabs\ = 2.9090 ---} If the central trough was fit with only one
line, the column density was found to be \lognhi\ $>$ 16.8. However,
there is no Lyman break feature around 3565~\AA. \lyb\ has an
asymmetrical profile. Therefore we fit the trough with two components
having \lognhi\ = 16.09 and 15.60.  The best-fitting model for \lya\
and \lyb\ is slightly inconsistent with \lyg\ and \lyd.

{\it \zabs\ = 3.2277 ---} This is a very weak system with \lognhi\ =
15.33, that may be a strong \lya\ forest member produced by an
intergalactic cloud. There are no metal lines in the system. The
spectrum has a narrow data defect at \delv\ = $-$500 to $-$400~\kms\
from the main component in \lya\ window.

{\it \zabs\ = 3.3212 ---} Burles, Kirkman, \& Tytler (1999) fit the
central trough with four components positioned at \delv\ = $-$98.6, 0,
$+$98, and $+$155~\kms\ from the main component. We also fit the
trough with four components positioned at \delv\ =$-$90, 0, $+$100,
and $+$150~\kms\ from the main component, which is in good agreement
with the results of Burles, Kirkman, \& Tytler (1999).  The Lyman
break around 3940~\AA\ suggests that this system has a column density
larger than \lognhi\ $>$ 16.6. This system is within 5000~\kms\ of the
quasar. A \ion{C}{4} complex at \zabs\ = 2.40 around 5260 -- 5275~\AA\
is blended with \lya\ lines of this system.

\item{\it Q0054-2824 (\zem\ = 3.616). ---} In this quasar, SSB found
strong \ion{Mg}{2} lines at \zabs\ = 1.3412 and 1.4398, and three
\ion{Si}{4} lines at \zabs\ = 3.2791, 3.5068 and 3.5800, but
associated \ion{C}{4} lines were not detected. The \ion{Si}{4} system
at \zabs\ = 3.5800 is known to be associated with the conspicuous LLS
at \zabs\ = 3.585. Our spectrum ranges from 4090~\AA\ to
6510~\AA. Both \lya\ and \lyb\ are detected at \zabs\ = 2.987 --
3.616.

{\it \zabs\ = 3.2370 ---} This is a less reliable system, because the
spectrum contains only three Lyman lines (\lya, \lyb, and \lyg), and
the S/N ratio is very low (S/N = 17 at \lya). If the central trough is
fit with only one component, the column density of the main component
is \lognhi\ = 16.6. The ratio of $N_{1}$ (the largest \hi\ column
density in the fit) to $N_{2}$ (the second largest \hi\ column density
in the fit) is then $N_{1}/N_{2}$ $>$ 200, which is rather large
compared with the usual value of $N_{1}/N_{2}$ $\sim$ 0.3.  Therefore
we used two components to fit the trough. We detected \ion{Si}{3} and
\ion{Si}{4} in this system.

{\it \zabs\ = 3.3123 ---} The model fit for \lya, \lyb, and \lyg\ is
inconsistent with \lyd, which may be due to the low S/N ratio of the
spectrum around the \lyd\ lines. There are no corresponding metal
lines, despite the large column density of the main component,
\lognhi\ $>$ 16.

{\it \zabs\ = 3.4488 ---} \hi\ lines of orders higher than \lyd\ are
located in the Lyman continuum of the LLS at \zabs\ = 3.58. There are
no corresponding metal lines at this redshift.

{\it \zabs\ = 3.5113 ---} SSB detected a tentative \ion{Si}{4} doublet
at \zabs\ = 3.507. Referring to that line, we found a \hi\ system with
\lognhi\ = 15.9 at \zabs\ = 3.511. There are seven unidentified narrow
lines with 7 $<$ $b$ $<$ 14~\kms\ at 5487 -- 5499~\AA.

{\it \zabs\ = 3.5805 ---} This LLS was reported in SBS. Since many
components are heavily blended with each other in the central trough,
the profile of higher order lines were used as a reference, and the
profile was fit with four components. The Lyman break of this system
is detected around 4180~\AA. This system has various metal lines,
including \ion{C}{2}, \ion{Si}{2}, and \ion{Si}{4}. We did not find
any velocity shift between low-ionization ions (\ion{C}{2} and
\ion{Si}{2}) and high-ionization ions (\ion{Si}{4}), though such
velocity shifts are often seen in DLA systems (e.g., Lu et al. 1996b;
Prochaska et al. 2001). This system is within 5000~\kms\ of the
quasar.

\item{\it Q0119+1432 (\zem\ = 2.870). ---} This quasar was discovered
during the course of the Hamburg/CfA Bright Quasar Survey (Hagen et
al. 1995; Dobrzycki et al. 1996). No detailed analysis of this quasar
has been published. Our spectrum ranges from 4090~\AA\ to
6510~\AA. Both \lya\ and \lyb\ are detected at \zabs\ = 2.120 --
2.870.

{\it \zabs\ = 2.4299 ---} Since this system is at low redshift
compared to other \hi\ systems in this study, the number density of
\lya\ forest lines is relatively small around this system. The
spectrum covers only \lya\ and \lyb. Nonetheless the model is reliable
thanks to the low number density of \lya\ forest lines. Only six \hi\
lines are detected within 1000~\kms\ of the main component.

{\it \zabs\ = 2.5688 ---} The central trough of this system has a
simple profile, but the profiles of higher-order lines suggest that
this system has a complex structure of narrow \hi\ components. We used
four components to fit the trough. We did not detect any metal lines
in the system.

{\it \zabs\ = 2.6632 ---} The large column density of the main
component, \lognhi\ = 19.37, results in strong Doppler wings at both
sides of the line. A Lyman-break feature is also detected around
3350~\AA. We detected low-ionization \ion{Si}{2} and \ion{Si}{3}
lines, while high-ionization lines such as \ion{Si}{4} and \ion{C}{4}
were not detected.

\item{\it HE0130-4021 (\zem\ = 3.030). ---} This quasar was discovered
by Osmer \& Smith (1976). Kirkman et al. (2000) obtained the spectrum
of the quasar with a total integration time of 22 hr, and found an LLS
at \zabs\ = 2.8 with low D/H abundance ratio, D/H =
$3.4\times10^{-5}$. The spectrum covers the range 3630~\AA\ to
6070~\AA. Both \lya\ and \lyb\ can be detected at \zabs\ = 2.539 --
3.030.

{\it \zabs\ = 2.8581 ---} Though the spectrum covers \lya, \lyb, \lyg,
and \lyd, they are located in a region with low S/N ratio, S/N $<
10$. The main component has a column density just over the limiting
value of \lognhi\ = 15. Nonetheless, this system is probably not a
normal \lya\ forest line, because it is accompanied by many metal
lines such as \ion{C}{4}, \ion{Si}{4}, \ion{Si}{3}, and \ion{Si}{2}.

\item{\it Q0241-0146 (\zem\ = 4.040). ---} The emission lines of the
quasar, such as \lya, \ion{O}{1}, \ion{C}{2}, \ion{Si}{4}, and
\ion{C}{4} , are known to be very broad and rounded. Storrie-Lombardi
et al. (1996) found a DLA system at \zabs\ = 2.86 with \lognhi\ = 19.8
and a metal line system with \ion{Mg}{2} and \ion{Fe}{2} lines at
\zabs\ = 1.435. Our spectrum ranges from 4490~\AA\ to 6900~\AA. Both
\lya\ and \lyb\ were detected at \zabs\ = 3.377 -- 4.040. However, no
\hi\ lines with \lognhi\ $>$ 15 were detected in our spectrum.

\item{\it Q0249-2212 (\zem\ = 3.197). ---} SSB found a very strong
\ion{C}{4} system at \zabs\ =3.1036 and weak \ion{C}{4} systems at
\zabs\ =2.6736 and 3.1294. The LLS detected at \zabs\ = 2.937 has no
associated metal lines. Our spectrum ranges from 3500~\AA\ to
5020~\AA. Both \lya\ and \lyb\ are detected at \zabs\ = 2.412 --
3.129.

{\it \zabs\ = 2.6745 ---} This system is a sub-DLA system with a
column density \lognhi\ = 19.0, resulting in strong Doppler wings;
however, the S/N ratio is very low (S/N = 15 at \lya). The Lyman break
of this system at $\lambda$ = 3349~\AA\ is not covered by our
spectrum.

{\it \zabs\ = 2.9401 ---} This system is also detected in a region of
low S/N ratio (S/N = 11 at \lya). The main component has \lognhi\ =
17.2. There is a tentative Lyman break feature around 3595~\AA. This
system, however, does not have any metal lines, in spite of the large
column density; this has already been noted by SSB.

\item{\it HE0322-3213 (\zem\ = 3.302). ---} Coordinate of this quasar
is given in Kirkman et al. (2005). Our spectrum ranges from 3830~\AA\
to 5350~\AA. Both \lya\ and \lyb\ are detected at \zabs\ = 2.734 --
3.317.

{\it \zabs\ = 3.0812 ---} We fit the central trough with one component
having \lognhi\ = 15.68. The lines around the main component have
column densities similar to that of the main component; \lognhi\ =
14.81, 14.19, 14.48, and 14.86 at \delv\ = $-$600, $+$250, $+$550, and
$+$900~\kms\ from the main component.

{\it \zabs\ = 3.1739 ---} Our spectrum covers from \lya\ to Ly11 of
this system. It was not possible to fit the central trough well, as it
is asymmetrical. This effect is probably artificial, because it seems
to be caused by the failure of continuum fitting, as is often the case
for the spectrum around strong absorption features. Nonetheless, our
fitting model is reliable to some extent, since the profiles of higher
orders are fit very well. This system is accompanied by five
\ion{Si}{2} lines.

{\it \zabs\ = 3.1960 ---} This system is only 1600~\kms\ redward of
the system at \zabs\ = 3.1739. Various metal lines, such as
\ion{C}{2}, \ion{Si}{2}, and \ion{Si}{3}, were detected in the
system. Though the spectrum has a wide data defect between 5126~\AA\
and 5132~\AA , corresponding to $\Delta v$ = 600 -- 900~\kms\ from the
main component, we were able to fit the regions with reference to the
profiles of higher orders.

{\it \zabs\ = 3.3169 ---} Our spectrum covers from \lya\ to Lyman
limit of this system, and the S/N ratio is very high (e.g., S/N = 103
at \lya). There are many broad and smooth components blueward of the
main component, while only narrow lines were detected in regions
redder than the main component. The narrow line clustering around
5262~\AA\ corresponds to unidentified metal lines. This system has
corresponding \ion{C}{2} and \ion{C}{3} lines.

\item{\it Q0336-0143 (\zem\ = 3.197). ---} This quasar was discovered
in the course of the Large Bright Quasar Survey (LBQS), and is known
to have a DLA system at \zabs\ = 3.061 with \lognhi\ = 21.18 (Lu et
al. 1993). \ion{Fe}{2}, \ion{Si}{2}, \ion{Si}{3}, \ion{Si}{4},
\ion{C}{2}, \ion{Al}{2}, and \ion{O}{1} are associated with this DLA
system. Lu et al. (1993) also detected an \ion{Na}{1} doublet at
\zabs\ = 0.1666, though this identification is less reliable because
of line blending with \ion{Al}{2}~$\lambda$1671 at \zabs\ =
3.1146. The \ion{Mg}{2} doublet at \zabs\ = 1.456 is also uncertain,
because the blue and red members of the doublet poorly agree in
redshift. The system also provides accurate measurements of uncommon
metal lines such as \ion{Ar}{1}, \ion{P}{2} and \ion{Ni}{2} (Prochaska
et al. 2001). Our spectrum ranges from 3940~\AA\ to 6390~\AA. Both
\lya\ and \lyb\ are detected at \zabs\ = 2.841 --
3.197. Unfortunately, the low S/N ratio of the spectrum prevents us
from detecting not only this DLA system but also other \hi\ systems
with \lognhi\ $>$ 15.

\item{\it Q0450-1310 (\zem\ = 2.300). ---} This quasar was discovered
by C. Hazard, and first studied by SBS and Steidel \& Sargent
(1992). They found one \ion{Fe}{2} line at \zabs\ = 1.1745, three
\ion{Mg}{2} lines at \zabs\ = 0.4940, 1.2291 and 1.3108, and three
\ion{C}{4} and \ion{Si}{4} lines at \zabs\ = 2.0669, 2.1063 and
2.2315. Petitjean et al. (1994) found an additional \ion{Mg}{2} line
at \zabs\ = 0.548, and four \ion{C}{4} lines at \zabs\ = 1.4422,
1.5223, 1.6967 and 1.9985. The system at \zabs\ = 2.0669 is a DLA
candidate, because it has a strong \lya\ line with large rest-frame
equivalent width ($W_{rest}$ = 6 \AA), and corresponding \ion{O}{1}
line which is often detected in DLA systems. The system at \zabs\ =
2.2315 is probably associated with the quasar, as the velocity
difference between the two is only 2080~\kms, and because the system
has a high-ionization \ion{N}{5} doublet which is usually detected in
the systems physically associated to quasars.  Our spectrum ranges
from 3390~\AA\ to 4910~\AA, and most of the region is redder than the
peak of the \lya\ emission lines at the redshift of the quasar, \zabs\
= 2.300.  Therefore we could not detect any \hi\ lines in our spectrum
with \lognhi\ $>$ 15.

\item{\it Q0636+6801 (\zem=3.178). ---} This quasar at \zem\ = 3.178
is one of the most luminous quasars known, and is listed as a radio
quasar in Hewitt \& Burbidge (1987). There are several \ion{C}{4}
absorption systems at \zabs\ = 2.4754, 2.8051, 2.9040, 3.0174, and
3.0589. The system at \zabs\ = 2.9040 is associated with the LLS at
\zabs\ = 2.909. At lower redshift, the \ion{Mg}{2} system is detected
at \zabs\ = 1.2941 (SSB). Our spectrum ranges from 3560~\AA\ to
6520~\AA. Both \lya\ and \lyb\ were detected at \zabs\ = 2.471 --
3.178.

{\it \zabs\ = 2.6825 ---} The absorption lines around this system were
well fit due to the high S/N ratio (S/N = 64 at \lya) and the low
number density of \lya\ forest lines around the system. The \lyg\
lines of the system are blanketed by the Lyman continuum of the LLS at
\zabs\ =2.904.

{\it \zabs\ = 2.8685 ---} The spectrum includes the \lya\ to Ly8 lines
of the system, though Ly7 and Ly8 are blanketed by the Lyman continuum
of the LLS at \zabs\ = 2.904. Four lines between \delv\ = 400~\kms\
and 700~\kms\ from the main component in \lya\ window are probably not
\hi\ lines, since corresponding lines of higher orders are not
detected.

{\it \zabs\ = 2.9039 ---} This system is an LLS with a large column
density, and has a clear Lyman break around 3570~\AA. Songaila \&
Cowie (1996) have already evaluated the column density of the system,
finding \lognhi\ = 17.8. Our best fit to this line gave \lognhi\ =
18.22. We detected 8 \ion{C}{4} and 6 \ion{Si}{4} doublets in the
system, though \ion{Si}{4}~$\lambda$1394 components are affected by
the spectrum gap. The system also has three \ion{O}{1} lines, which
strongly suggests that it is in a low ionized state, surrounded by gas
clouds of large column density.

{\it \zabs\ = 3.0135 ---} This system, with \lognhi\ = 15.79, has two
\ion{C}{4} doublets. The spectrum ranges from \lya\ to the Lyman limit
of the system. Misawa et al. (2007) identified this system as a quasar
intrinsic system, based on the partial coverage analysis of the
\ion{C}{4} doublet.

{\it \zabs\ = 3.0675 ---} This is a weak system with \lognhi\ =
15.28. The fitting model is very reliable because the spectrum ranges
from \lya\ to the Lyman limit of the system with a high S/N ratio
(e.g., S/N = 117 at \lya, 44 at \lyb, and 13 at Ly10).

\item{\it Q0642+4454 (\zem=3.408). ---} This quasar is one of the
quasars for which LLS absorption was detected for the first time
(Carswell et al. 1975). However, we did not detect the LLS which had
been discovered at \zabs\ = 3.295 by Carswell et al. (1975). SSB found
three \ion{C}{4} systems at \zabs\ = 2.9724, 3.1238, and 3.2483, and
one \ion{Mg}{2} system at \zabs\ = 1.2464. Our spectrum ranges from
3930~\AA\ to 6380~\AA. Both \lya\ and \lyb\ are detected at \zabs\ =
2.831 -- 3.408.

{\it \zabs\ = 2.9726 ---} The model is less reliable here, because
only \lya\ and \lyb\ could be used as reference lines. Nonetheless,
this system would be expected to have a large column density, as there
are various metal lines such as \ion{C}{4}, \ion{Si}{2}, and
\ion{Si}{4}. We fit the central trough with one component having
\lognhi\ = 17.36; however, the ratio of $N_{1}$ to $N_{2}$ is too
large, $N_{1}/N_{2}$ $>$ $10^{3}$. This component may be resolved into
multiple narrow components. This system was classified into a quasar
intrinsic system (Misawa et al. 2007).

{\it \zabs\ = 3.1230 ---} This system is a sub-DLA with \lognhi\ =
19.48. The existence of \ion{O}{1} line in the system strongly
suggests that the system has large column density, because \ion{O}{1}
lines are rarely seen except in DLAs. The Lyman break of the DLA
system is seen around 3750~\AA\ in the low resolution spectrum in
SSB. We did not detect \ion{C}{4} lines in this system, though
\ion{Si}{4} and \ion{C}{2} lines were detected.

{\it \zabs\ = 3.1922 ---} This weak system does not have any
corresponding metal lines. Around \delv\ = 700~\kms\ from the main
component, there are two unidentified metal lines. The sharp spike at
\delv\ = $-$150~\kms\ from the main component is a data defect.

{\it \zabs\ = 3.2290 ---} There are two components with very similar
column densities, \lognhi\ = 15.52 and 15.37, at 5145~\AA\ and
5150~\AA. The ratio of $N_{1}$ to $N_{2}$ is near unity. This system
is only 1300~\kms\ blueward of the system at \zabs\ = 3.248.

{\it \zabs\ = 3.2476 ---} Based on the \ion{C}{4} doublet at \zabs\ =
3.248 detected by SSB, we found a corresponding \hi\ line with
\lognhi\ $>$ 15. Though the central trough is damaged by a wide data
defect of 2~\AA\ width, it was possible to fit the trough using the
features of higher orders.  The most interesting feature of this
system is that the high-ionization lines (e.g., \ion{Si}{4}) are
surrounded by the low-ionization lines (e.g., \ion{C}{2} and
\ion{C}{3}), which is the reverse of the trend usually seen in DLA
systems.

{\it \zabs\ = 3.3427 ---} Since the \lya\ lines of this system are
also affected by the 2~\AA-wide data defect, we fit the system by
referring to the profiles of higher orders. The system is within
5000~\kms\ of the quasar.

\item{\it HS0757+5218 (\zem\ = 3.240). ---} This quasar was discovered
during the course of the Hamburg/CfA Bright Quasar Survey. No detailed
analysis of the quasar spectrum has been published. Our spectrum
ranges from 3590~\AA\ to 5120~\AA.  Both \lya\ and \lyb\ are detected
at \zabs\ = 2.500 -- 3.212.

{\it \zabs\ = 2.7261 ---} Our spectrum covers the \lya, \lyb, and
\lyg\ lines of the system, though \lyg\ is not useful because of low
S/N ratio. Four narrow components were detected at 600~\kms\ --
800~\kms\ from \lya\ of the main component. They are unidentified
metal lines.

{\it \zabs\ = 2.8922 ---} The central trough has a wide and smooth
profile. We fit it with a single component having \lognhi\ =
18.34. However, the Doppler parameter of the component is too large,
$b$ = 54~\kms. This component may be resolved into multiple narrow
components.

{\it \zabs\ = 3.0398 ---} Since this sub-DLA system has strong damping
wings, the spectrum could not be normalized correctly around the \lya\
line, as is often the case for echelle-formatted spectra. Nonetheless,
we fit the strongly damped feature with five components by referring
to higher orders. Metal lines corresponding to this system are not
detected, despite the large column density of the system.

\item{\it Q0805+0441 (\zem\ = 2.880). ---} This quasar is well known
as the radio source 4C 05.34. Chen et al. (1981) first studied the
absorption systems of the quasar in detail. SSB found three \ion{C}{4}
systems at \zabs\ = 2.4509, 2.4742, and 2.8758, and one \ion{Mg}{2}
system at \zabs\ = 0.9598. There is also an LLS at \zabs\ = 2.651, but
this system does not have associated heavy element lines. Our spectrum
ranges from 3800~\AA\ to 6190~\AA. Both \lya\ and \lyb\ are detected
at \zabs\ = 2.705 -- 2.880.

{\it \zabs\ = 2.7719 ---} There is a weak upward spike at the center
of the \lyb\ line in the central trough. Based on this feature, we fit
the trough with two components having \lognhi\ = 16.30 and 15.14. We
did not detect any metal absorption lines.

{\it \zabs\ = 2.8113 ---} Because the central trough has an
asymmetrical feature, we used two components to fit the profile. The
narrow line at 4625~\AA\ was identified as \ion{Si}{2}~$\lambda$1193
by Chen et al. (1981). However, we identify it as
\ion{Mg}{2}~$\lambda$2796 at \zabs\ = 0.654, as the corresponding
\ion{Mg}{2} line of this doublet, \ion{Mg}{2}~$\lambda$2803, is
detected at 4627~\AA\ in our spectrum.

\item{\it Q0831+1248 (\zem\ = 2.734). ---} SBS found an absorption
system at \zabs\ = 2.0844 with a \ion{C}{4} doublet and
\ion{Al}{2}~$\lambda$1671 line. Lanzetta et al. (1991) found another
system at \zabs\ = 2.796 with \ion{Si}{2} and \ion{C}{2} lines. Our
spectrum covers from 3790~\AA\ to 6190~\AA. Both \lya\ and \lyb\ are
detected at \zabs\ = 2.695 -- 2.734.

{\it \zabs\ = 2.7300 ---} Metal lines such as \ion{C}{4} and
\ion{Si}{4} were detected in this system. As the system is just
blueward of the emission redshift of the quasar at \zem\ = 2.734,
there are few absorption features redward of this system. This system
may be intrinsically associated with the quasar itself, as the
velocity distance from the quasar is small.

\item{\it HE0940-1050 (\zem\ = 3.080). ---} This quasar was discovered
during the course of the Hamburg/CfA Bright Quasar Survey. Reimers et
al. (1995) detected four heavy-element systems at \zabs\ = 2.82
(\ion{C}{4}), 2.32 (\ion{C}{4}), 1.918 (\ion{Fe}{2}, \ion{Al}{2}, and
\ion{Si}{2}), and 1.06 (\ion{Mg}{2}) with the 4~\AA\ resolution
spectrum taken with the ESO 1.5m telescope.  Reimers et al. (1995) did
not detect any flux from the quasar below 3200\AA\ in a spectrum from
IUE, probably due to a Lyman limit systems at \zabs\ $\sim$ 2.82
and/or 2.32.  Our spectrum ranges from 3610~\AA\ to 6030~\AA.  Both
\lya\ and \lyb\ are detected at \zabs\ = 2.519 -- 3.080.

{\it \zabs\ = 2.8283 ---} Our spectrum covers \lya\ to \lyd\ in this
system. However, \lyg\ and \lyd\ are in low S/N regions. Three strong
\hi\ components with \lognhi\ = 15.9, 16.4, and 16.0 are located at
\delv\ = $-$300, 0, and $+$490~\kms\ from the main component. Seven
\ion{C}{4} doublets and five \ion{Si}{4} doublets, \ion{C}{2} and
\ion{Si}{3} lines were detected in the system.

{\it \zabs\ = 2.8610 ---} We fit the central trough with one component
having \lognhi\ = 17.06. However, this component may be resolved into
multiple narrower components, because the ratio of $N_{1}$ to $N_{2}$
is unusually large, $N_{1}/N_{2}$ $\sim$ 300. We also detected
\ion{Si}{4} and \ion{C}{4} doublets in the system.

{\it \zabs\ = 2.9174 ---} We fit the main trough with five components,
referring not only to the profiles of higher orders but also to the
weak upward spikes in the damped region of the \lya\ profile. We also
detected \ion{Si}{3} and \ion{Si}{4} lines at the redshift of the main
component.

{\it \zabs\ = 3.0387 ---} This system is within 5000~\kms\ of the
emission redshift of the quasar. At the bluer region of the main
component there are several unidentified metal lines.  They could be
identified as \ion{C}{4} lines at \zabs\ $\sim$ 2.17, but one of them,
at \zabs\ = 2.158, has a Doppler parameter of $b$ = 28~\kms\ which is
unusually large for an ordinary \ion{C}{4} line.

\item{\it Q1009+2956 (\zem=2.644). ---} Burles \& Tytler (1998b)
presented a measurement of the D/H ratio in the metal-poor absorption
system at \zabs\ = 2.504. They estimated the D/H ratio in the system
to be $\log$(D/H) = $-4.40^{+0.06}_{-0.08}$ at the 67\% confidence
level. Our spectrum ranges from 3090~\AA\ to 4620~\AA. Both \lya\ and
\lyb\ are detected at \zabs\ = 2.013 -- 2.644.

{\it \zabs\ = 2.1432 ---} If the main trough is fit with a single
component, it is found to be very broad, $b \sim$ 40~\kms. We
therefore used three components with narrower profiles, $b$ = 26, 28,
and 31~\kms. The fitting model is still uncertain, however, as we
could refer to only \lya\ and \lyb , and both of them are entirely
damped.

{\it \zabs\ = 2.4069 ---} We referred to \lya, \lyb, \lyg, and \lyd\
for the profile fitting. The lines of higher orders are blanketed by
the Lyman continuum of the LLS at \zabs\ = 2.50. We fit the central
trough with one component having \lognhi\ = 18.8, and $b$ =
48~\kms. Although this component, with its rather large $b$ value, may
be resolved into several narrow components, there is too little
information to be able to separate them out. This system is
accompanied by low-ionization metal lines such as \ion{O}{1},
\ion{C}{2}, \ion{Si}{2}, and \ion{Si}{3}.

{\it \zabs\ = 2.4292 ---} This system is only 1000~\kms\ to the red of
the system at \zabs\ = 2.407. The spectrum covers the system from
\lya\ to Ly6, but they are all damped. We fit the central trough with
one component having \lognhi\ = 17.43. However, the ratio of $N_{1}$
to $N_{2}$ is unusually large, $\sim$ 350. The main component may be
resolved into multiple narrower components. We also detected
low-ionization metal lines such as \ion{C}{2}, \ion{Si}{2}, and
\ion{Si}{3}.

{\it \zabs\ = 2.5037 ---} This system is known to contain a candidate
\ion{D}{1} line. Detailed fitting was applied by Burles \& Tytler
(1998b), and they measured the D/H ratio to be $\log$(D/H) =
$-4.40^{+0.06}_{-0.08}$. The \hi\ column density of the system was
estimated by Burles \& Tytler (1998b) to be \lognhi\ = 17.35, which is
similar to our result, \lognhi\ = 17.26. The \ion{D}{1} candidate is
also detected in our spectrum. Additionally, we detect \ion{Si}{3}
lines.

\item{\it Q1017+1055 (\zem\ = 3.156). ---} SBS found the BAL type
features for \ion{C}{4} and \ion{Si}{4} lines at \zabs\ = 2.9720,
though there were no BAL type \lya\ absorption lines. Three \ion{C}{4}
systems at \zabs\ = 2.5401, 2.9970, and 3.1101, two \ion{Mg}{2}
systems at \zabs\ = 0.974 and 1.2401, and an LLS at \zabs\ = 3.048
were detected. Our spectrum ranges from 3890~\AA\ to 6300~\AA. Both
\lya\ and \lyb\ are detected at \zabs\ = 2.792 -- 3.156.

{\it \zabs\ = 2.9403 ---} This system is detected based on only \lya\
and \lyb\ lines in the regions with a low S/N ratio (S/N = 12 at \lya\
and 8.2 at \lyb). No metal lines were detected in the system.

{\it \zabs\ = 3.0096 ---} This system is probably identical to the
\ion{C}{4} system detected at \zabs\ = 2.997 in SBS. We detected
\ion{C}{4} and \ion{Si}{4} doublets in this system, as previously
described in SBS.

{\it \zabs\ = 3.0548 ---} The central trough was fit with four
components having column densities \lognhi\ = 15.4, 17.1, 14.9, and
14.0. This model, however, is somewhat uncertain, as the S/N ratio is
very low (S/N = 25 at \lya\ and 8.9 at \lyb). The system is
accompanied by three weak \ion{Si}{4} doublets.

{\it \zabs\ = 3.1120 ---} This system is detected based on the
\ion{C}{4} doublet at \zabs\ = 3.11 described in SBS, though the
\ion{C}{4} doublet itself is not covered by our spectrum. We detect
two \ion{Si}{4} doublets in the system. Due to the asymmetrical
feature present in the \lyd\ profile of the central trough, we fit it
with two components having \lognhi\ = 15.3 and 15.0. This system is
within 5000~\kms\ of the quasar.

\item{\it Q1055+4611 (\zabs\ = 4.118). ---} The system at \zabs\ =
3.32 has been identified as a DLA with \lognhi\ = 20.34 (Lu, Sargent
\& Barlow 1998). Storrie-Lombardi \& Wolfe (2000) also found another
DLA system at \zabs\ = 3.05, and estimated the column density of the
DLA to be \lognhi\ = 20.3; this was confirmed by P\'eroux et
al. (2001). There is also an LLS at \zabs\ = 2.90 with large optical
depth, $\tau$ = 4.4 (P\'eroux et al. 2001). Our spectrum covers a
range of 4450~\AA\ to 6900~\AA. Both \lya\ and \lyb\ are detected at
\zabs\ = 3.338 -- 4.118.

{\it \zabs\ = 3.8252 ---} The central trough was fit with three
components with \lognhi\ = 14.9, 16.0, and 15.6 at \delv\ = $-$65, 0,
and 100~\kms\ from the main component. There may exist additional
components between \delv\ = 0 and $+$100~\kms, because there is an
excessive residual flux at \lyg\ and \lyd. The system is accompanied
by three \ion{Si}{4} doublets. The \hi\ lines of orders higher than
\lyd\ are all blanketed by the Lyman continuum of the LLS at \zabs\ =
3.93.

{\it \zabs\ = 3.8495 ---} This system is shifted redward of the system
at \zabs\ = 3.82 by just 1000~\kms . We fit the central trough with
two components, using the profiles of \lyd, \lye, and Ly8 as a
reference. The system contains \ion{Si}{3} and \ion{Si}{4} lines. The
Lyman break of the system is not seen, as it is blanketed by the Lyman
continuum of the LLS at \zabs\ = 3.93.

{\it \zabs\ = 3.9343 ---} The optical depth of the system at the Lyman
break was estimated to be $\tau$ $>$ 3, which corresponds to a column
density of \lognhi\ $>$ 17.7. Using this result as a reference, we fit
the main trough with two components having a total column density of
\lognhi\ = 17.34. Although there is a data defect at \delv\ = $-$750
-- $-$650~\kms\ from the main component in \lya\ window, the lines
there can be fit using the profiles of higher orders as a reference.

\item{\it HS1103+6416 (\zem\ = 2.191). ---} This quasar was discovered
during the Hamburg Quasar Survey (Hagen et al. 1995). K\"ohler et
al. (1999) have studied the complex LLS at \zabs\ = 1.892, using both
ultraviolet spectra taken with the HST and optical high resolution
spectra taken with Keck. They found that the complex absorption lines
are distributed with a velocity width of $\sim$ 200~\kms; they also
found the system to contain at least 11 narrow components with various
ionization levels. Our spectrum ranges from 3180~\AA\ to
5790~\AA. Both \lya\ and \lyb\ are detected at \zabs\ = 2.100 --
2.191.

\item{\it Q1107+4847 (\zem\ = 3.000). ---} Carballo et al. (1995)
first observed this quasar with a moderate resolution of 40 --
120~\kms, and detected three \ion{C}{4} systems at \zabs\ = 2.697,
2.724, and 2.760. However, two of them, at \zabs\ = 2.697 and 2.724,
are doubtful, because the \ion{C}{4}~$\lambda$1551 components are
absent. Our spectrum ranges from 3730~\AA\ to 6170~\AA. Both \lya\ and
\lyb\ are detected at \zabs\ = 2.636 -- 3.000.

{\it \zabs\ = 2.7243 ---} The main component has a rather large
Doppler parameter, $b$ = 48~\kms. However, we cannot resolve the
component into narrower components, as our spectrum covers only \lya\
and \lyb\ lines, which are both strongly damped. This system has four
\ion{C}{4} and three \ion{Si}{4} doublets. There are many unidentified
metal lines at \delv\ = 700 -- 1000~\kms\ from the main component in
\lya\ window. Misawa et al. (2007) identified this system as a quasar
intrinsic system.

{\it \zabs\ = 2.7629 ---} This partial DLA system, with column density
\lognhi\ = 19.13, is accompanied by various metal lines such as
\ion{C}{4}, \ion{Si}{2}, \ion{Si}{3}, \ion{Si}{4}, and
\ion{O}{1}. These metal lines result in complex structure, as
described by Carballo et al. (1995). Our spectrum covers only \lya\
and \lyb\ lines, which are both strongly damped. Therefore we used the
velocity distribution of metal lines as a reference for fitting the
\hi\ components in the central trough.

{\it \zabs\ = 2.8703 ---} Based on the profiles of \lya, \lyb, and
\lyg\ lines, it was possible to separate the components in the central
trough sufficiently. A weak \ion{Si}{4} doublet is also detected in
the system. However, we did not detect the corresponding \ion{C}{4}
doublet, because it is located at the region of the spectral gap.

\item{\it Q1157+3143 (\zem\ = 2.992). ---} There are two LLS
candidates at \zabs\ $\sim$ 2.94 and 2.77 in the spectrum of this
quasar (Kirkman \& Tytler 1999). However, the \hi\ column density of
the LLS at \zabs\ $\sim$ 2.77 has not been evaluated exactly, because
the LLS at \zabs\ $\sim$ 2.94 blots out the spectrum below 3600 \AA,
which prevents the detection of \hi\ lines of orders higher than \lyg\
for the LLS at \zabs\ $\sim$ 2.77. Nonetheless, the system probably
has a large column density because the system is accompanied not only
by high ionization ions such as \ion{Si}{4}, \ion{C}{4} and
\ion{O}{6}, but also by low ionization ions such as \ion{C}{2} and
\ion{Si}{2}. These low ionization lines are expected to be present in
systems with large column densities. Our spectrum covers from
3790~\AA\ to 6190~\AA. Both \lya\ and \lyb\ are detected at \zabs\ =
2.695 -- 2.992.

{\it \zabs\ = 2.7710 ---} This is one of the two LLSs detected in the
direction of this quasar by Kirkman \& Tytler (1999). As our spectrum
covers only \lya\ and \lyb\ and both are damped, we fit the central
trough with one component having \lognhi\ = 17.63. The system has 7
\ion{C}{4}, 5 \ion{Si}{4}, 2 \ion{C}{2}, and 6 \ion{Si}{3} lines. The
Lyman break is blanketed by the Lyman continuum region of another LLS
at \zabs\ $\sim$ 2.94.

{\it \zabs\ = 2.8757 ---} Around the center of this system there are
two \hi\ lines with very similar column densities, \lognhi\ = 15.54
and 15.66, and with a velocity separation of 210~\kms. Interestingly,
only one of them is accompanied by various metal lines, such as
\ion{Si}{3}, \ion{Si}{4}, and \ion{C}{4}.

{\it \zabs\ = 2.9437 ---} This system is another LLS detected by
Kirkman \& Tytler (1999) with \lognhi\ = 17.44. This LLS has a
companion \hi\ line with similar column density of \lognhi\ = 17.16
having a separation of 250~\kms. \ion{Si}{4}, and \ion{Si}{2} lines
were detected in the system. Two \ion{O}{1} lines were also
detected. However, the detection of \ion{O}{1} lines is tentative,
because they have shifted as much as 400~\kms\ blueward of the main
component.

\item{\it Q1208+1011 (\zem\ = 3.803). ---} This is a candidate
gravitationally lensed quasar (Bahcall et al. 1992). Two point sources
with magnitudes of $V$ = 18.3 and $V$ = 19.8 are separated by
0$^{\prime\prime}$.47. The spectra of both objects have strong
\lya$+$\ion{N}{5} and \ion{O}{6}$+$\lyb\ emission lines at \zem\
$\sim$ 3.8, and many common absorption features. The low resolution
(25~\AA) spectrum of Bahcall et al. (1992), however, could not resolve
whether these objects are (i) gravitational lensed images, or (ii) two
quasars with a projected separation of a few kpc. Our spectrum ranges
from 3730~\AA\ to 6170~\AA. Both \lya\ and \lyb\ were detected at
\zabs\ = 2.636 -- 3.803.

{\it \zabs\ = 3.3846 ---} The spectrum covers all Lyman series orders
of this system, but they are positioned in regions with low S/N ratio
(e.g., S/N = 24 at \lya). As the central trough is not separated into
narrow components up to Ly10, we fit the trough with a single
component having \lognhi\ = 17.35. We did not detect any metal lines
in the system.

{\it \zabs\ = 3.4596 ---} The \lya\ lines of this system are blended
with \ion{C}{4} lines at \zabs\ $\sim$ 2.50. Our fitting model, when
applied to the \lya, \lyg, and \lyd\ lines, does not agree with the
wings of \lyb, which suggests that the wings are not \lyb\ lines at
\zabs\ $\sim$ 3.46 but other, lower-redshift lines such as \lya .

{\it \zabs\ = 3.5195 ---} We fit the central trough with two
components based on the features of Ly8 and Ly10. The profiles of
\lya\ -- Ly7 are all saturated.

{\it \zabs\ = 3.7206 ---} We fit the system with a component with a
rather small column density, \lognhi\ = 15.48. No metal lines
associated with the system were detected in our spectrum.

\item{\it Q1244+1129 (\zem\ = 2.960). ---} This quasar was discovered
in the Hamburg/CfA Bright Quasar Survey.  No results of detailed
analysis for this quasar have been published.  Our spectrum ranges
from 3370~\AA\ to 4880~\AA. Both \lya\ and \lyb\ were detected at
\zabs\ = 2.285 -- 2.960.

{\it \zabs\ = 2.9318 ---} The central trough, which has an
asymmetrical profile, was fit with four components using higher order
\hi\ lines as a reference. No metal lines were detected in the
system. The system is within 5000~\kms\ of the quasar.

\item{\it Q1251+3644 (\zem\ = 2.988). ---} There is a Lyman break at
3303~\AA\ in this system, produced by the LLS at \zabs\ = 2.614
(Stengler-Larrea et al. 1995). Our spectrum ranges from 3790~\AA\ to
6190~\AA. Both \lya\ and \lyb\ are detected at \zabs\ = 2.695 --
2.988.

{\it \zabs\ = 2.8684 ---} Only \lya\ and \lyb\ lines are covered by
the spectrum. Although both orders are affected by several data
defects, the fit model is good. We did not detect any metal lines in
this system.

\item{\it Q1330+0108 (\zem\ = 3.510). ---} A Lyman break at 4034~\AA,
produced by the LLS absorber at \zabs\ = 3.414, was detected
(Stengler-Larrea et al. 1995). Our spectrum ranges from 4030~\AA\ to
6450~\AA. Both \lya\ and \lyb\ are detected at \zabs\ = 2.929 --
3.510.

\item{\it Q1334-0033 (\zem\ = 2.801). ---} Although Wolfe et
al. (1995) attempted to find DLA systems in the spectrum of this
quasar, they found no candidates for DLA systems in the observed
redshift range, 1.634 $<$ $z$ $<$ 2.745. No absorption systems in the
spectrum of this LBQS quasar have been published. Our spectrum ranges
from 3730~\AA\ to 6170~\AA. Both \lya\ and \lyb\ are detected at
\zabs\ = 2.636 -- 2.801.

{\it \zabs\ = 2.7572 ---} Our spectrum covers only \lya\ and \lyb\
lines, and the S/N ratio around \lyb\ is very low, S/N $<$
10. Therefore the fitting model is not good. However, the existence of
three \ion{C}{4} doublets in the system strongly suggests that the
system is real. This system is within 5000~\kms\ of the quasar.

\item{\it Q1337+2832 (\zem\ = 2.537). ---} This quasar was discovered
by the CFHT/MMT survey (Crampton et al. 1988,1989). Our spectrum
covers a range of 3170~\AA\ to 4710~\AA. Both \lya\ and \lyb\ are
detected at \zabs\ = 2.091 -- 2.537.

{\it \zabs\ = 2.4336 ---} The optical depth of the system at the Lyman
break was estimated to be $\tau$ $>$ 3, corresponding to a column
density of \lognhi\ $>$ 17.7. In our spectrum, we used only lines of
orders up to \lyd, because the S/N ratios of the higher orders are
very low (S/N = 2.5 at \lye). We fit the central trough with four
components having a total column density of \lognhi\ = 18.9, based on
the profile of \lyd. While high ionization lines such as the
\ion{C}{4} and \ion{Si}{4} doublets of this system are not covered by
the spectrum, low ionization lines (e.g., \ion{C}{2}, \ion{Si}{2},
\ion{Si}{3}, and \ion{O}{1}) were detected.

{\it \zabs\ = 2.5228 ---} This system is within 5000~\kms\ of the
quasar. The central trough was fit with two components having
column densities of \lognhi\ = 15.81 and 14.44, because the trough has
an asymmetrical profile at \lya\ and \lyb. We did not detect any metal
lines in the system.

\item{\it Q1422+2309 (\zem\ = 3.611). ---} Kirkman \& Tytler (1997b)
have detected the LLS at \zabs\ = 3.3816 with high ionization metal
lines such as \ion{C}{4} and \ion{O}{6}. The \hi\ column density of
the LLS was estimated to be \lognhi\ = 20.6 assuming the LLS is
collisionally ionized, and \lognhi\ = 19.9 if it is photoionized. Our
spectrum ranges from 3740~\AA\ to 6180~\AA. Both \lya\ and \lyb\ are
detected at \zabs\ = 2.646 -- 3.611.

{\it \zabs\ = 3.3825 ---} The spectrum covers the wavelength range
from \lya\ to the Lyman limit of this system. The S/N ratio is high
over all orders. We fit the central trough with two components having
similar column densities, \lognhi\ = 16.5 and 16.3, using the profile
of orders higher than \lye\ as a reference.

{\it \zabs\ = 3.5362 ---} The wavelength range of our spectrum also
covers all Lyman orders. As the \lya\ lines of the system are located
in the gap of the spectrum, we fit them by reference to the higher
orders. We resolved the central trough into 6 narrow components with
column densities of \lognhi\ = 14.5 -- 15.9. Three \ion{C}{2} and four
\ion{Si}{3} lines were also detected. This system is within 5000~\kms\
of the quasar.

\item{\it Q1425+6039 (\zem\ = 3.165). ---} This quasar is one of four
very luminous quasars at \zem\ $>$ 2 found by the Second Quasar Survey
(Stepanian et al. 1990 and references therein). Stepanian et
al. (1996) carried out follow-up spectroscopy for this quasar with the
6m telescope of the Special Astrophysical Observatory, and detected a
strong DLA system with \lognhi\ = 20.4 at \zabs\ = 2.826. This system
contains not only low-ionization ions (e.g., \ion{C}{2}, \ion{Si}{2},
\ion{Fe}{2}, and \ion{Al}{2}), but also high-ionization ions (e.g.,
\ion{O}{6}, \ion{N}{5}, \ion{Si}{4}, and \ion{C}{4}). Our spectrum
ranges from 3730~\AA\ to 6170~\AA. Both \lya\ and \lyb\ are detected
at \zabs\ = 2.636 -- 3.165.

{\it \zabs\ = 2.7700 ---} This partial DLA system has a large column
density of \lognhi\ = 19.4, and results in strong damping wings,
preventing us from detecting the weak components around the DLA
system. We fit the central trough with seven components. Three \hi\
components near the center of the system are located at the same
redshift as three \ion{Si}{3} components in the system. We fit the
Doppler wings with four components using the profiles of \lyb\ lines
as a reference.  There are also \ion{Si}{2}, \ion{Si}{4}, and
\ion{C}{4} lines in the system. Although this system was classified as
a possible candidate for a quasar intrinsic system, it was less
reliable because of this DLA-like structure (Misawa et al. 2007).

{\it \zabs\ = 2.8258 ---} This DLA system has a large column density
of \lognhi\ = 20.0, which results in strong damping wings. Almost all
of the \hi\ lines within 1000~\kms\ of the main component are
blanketed by the wings. This system is not suitable for our study of
the structure of \hi\ systems. In this DLA system, 7 \ion{C}{4}, 7
\ion{C}{2}, 9 \ion{Si}{4}, 2 \ion{Si}{3} and 6 \ion{Si}{2} lines were
detected.

{\it \zabs\ = 3.0671 ---} We could not detect the Lyman break of the
system around 3740~\AA\ due to a low S/N ratio. The \lyd\ profile has
two upward spikes on the bottom of the central trough. If the trough
is fit with three narrow components based on these spikes, the Doppler
parameters of them are found to be $b \sim$ 7~\kms, which is too low
for \hi\ lines. Therefore we fit the trough with two components, using
the asymmetrical \lya\ profile as a reference.

{\it \zabs\ = 3.1356 ---} The central trough was fit with two
components having \lognhi\ = 16.7 and 16.2, using the profiles of
orders higher than \lye\ as a reference. The region between \delv\ =
$-$500 and $-$300~\kms\ from the main component in \lya\ window is
affected by the gap of the spectrum. Therefore we fit the region based
on lines of order higher than \lyb. The system is within 5000~\kms\ of
the emission redshift of the quasar.

\item{\it Q1442+2931 (\zem\ = 2.670). ---} Carballo et al. (1995)
found three \ion{C}{4} systems at \zabs\ = 2.330, 2.439, and 2.474 in
the spectrum of this quasar. The strongest system at \zabs\ = 2.439
was found to be accompanied by low-ionization lines such as \ion{O}{6}
and \ion{C}{2}. Carballo et al. (1995) also detected two \lya\
absorption lines with \lognhi\ $>$ 16 at \zabs\ = 2.555 and 2.617,
though no metal lines were detected in their redshift range. Our
spectrum ranges from 3740~\AA\ to 6180~\AA. Both \lya\ and \lyb\ are
detected at \zabs\ = 2.646 -- 2.670.

\item{\it Q1526+6701 (\zem\ = 3.020). ---} This quasar was discovered
during observations taken with the NRAO Green Bank 300 foot telescope
(Becker et al. 1991). Storrie-Lombardi \& Wolfe (2000) confirmed that
there were no candidate DLA systems in their observed redshift range,
1.955 $<$ $z$ $<$ 2.980. Our spectrum ranges from 3460~\AA\ to
4980~\AA. Both \lya\ and \lyb\ are detected at \zabs\ = 2.373 --
3.020.

{\it \zabs\ = 2.9751 ---} We detected the system in a region of low
S/N ratio (e.g., S/N = 24 at \lya, and 7.5 at \lyb). We fit the
central trough with three components using the profile of \lyb\ as a
reference. No metal lines were detected in the system. The system is
within 5000~\kms\ of the quasar.

\item{\it Q1548+0917 (\zem\ = 2.749). ---} SBS found a fairly weak,
but unambiguous, \ion{Mg}{2} doublet at \zabs\ = 0.7708 in the
spectrum of the quasar. They also detected three \ion{C}{4} doublets
at \zabs\ = 2.2484, 2.3195, and 2.4915. Our spectrum ranges from
3730~\AA\ to 6180~\AA, which does not cover the heavy element systems
detected in SBS.

\item{\it Q1554+3749 (\zem\ = 2.664). ---} This quasar was discovered
by the Palomar Transit Grism Survey (PTGS; Schneider, Schmidt, \& Gunn
1994). No absorption systems in the quasar have been published. Our
spectrum ranges from 3240~\AA\ to 4770~\AA. Both \lya\ and \lyb\ are
detected at \zabs\ = 2.159 -- 2.664.

{\it \zabs\ = 2.6127 ---} The central trough was fit with two
components having \lognhi\ = 18.0 and 14.5. Although the trough may be
separated into more than two components, we cannot resolve it with our
low S/N spectrum. The strong lines at \delv\ = $-$450~\kms\ and
$-$250~\kms\ from \lya\ of the main component are not \lya\ lines,
because there are no corresponding \lyb\ lines. They are also not
\lyb\ lines at higher redshift, since the corresponding redshift,
\zabs\ $\sim$ 3.27, is higher than the emission redshift of the
quasar, \zem\ = 2.664. Therefore they are likely to be very strong
metal absorption lines.

\item{\it HS1700+6416 (\zem\ = 2.722). ---} This quasar has been well
studied by both ground and space telescopes (Reimers et al. 1989; Sanz
et al. 1993; Reimers et al. 1992; Rodr\'iguez-Pascual et al. 1995), as
6 (sub-)LLS candidates have been identified at \zabs\ = 2.4336,
2.1681, 1.8465, 1.725, 1.1572 and 0.8642 with 16.0 $<$ \lognhi\ $<$
18.3. There are also 4 \ion{C}{4} doublets at \zabs\ = 2.7444, 2.7102,
2.5784 and 2.308. One of them is at \zabs\ $>$ \zem. Our spectrum
ranges from 3730~\AA\ to 6180~\AA. Both \lya\ and \lyb\ are detected
at \zabs\ = 2.636 -- 2.722. We detected the \ion{C}{4} systems at
\zabs\ = 2.7444 and 2.7102 in our spectrum. However, the corresponding
\hi\ lines are very weak, \lognhi\ $\sim$ 14.0.

\item{\it Q1759+7529 (\zem\ = 3.050). ---} Outram et al. (1999)
studied this quasar in detail. They detected a DLA system at \zabs\ =
2.625 with \lognhi\ = 20.76. We also found a partial DLA system at
\zabs\ = 2.910 with \lognhi\ = 19.80, which is accompanied by an LLS
with \lognhi\ = 17.02 having a separation of 420~\kms. In the spectrum
of the quasar, 9 \ion{C}{4} systems were detected at \zabs\ = 1.8848,
1.935, 2.4390, 2.484, 2.7871, 2.795, 2.835, 2.84, and 2.896. Three
systems at \zabs\ = 1.935, 2.484 and 2.625 have complex structures
with both low- and high-ionization lines. Galactic absorption lines of
\ion{Na}{1}~$\lambda\lambda$5892,5898 were also observed. Our spectrum
ranges from 3580~\AA\ to 5050~\AA. Both \lya\ and \lyb\ are detected
at \zabs\ = 2.490 -- 3.050.

{\it \zabs\ = 2.7953 ---} We fit the central trough with four
components having column densities of \lognhi\ = 13.4, 14.0, 15.3, and
14.9, based on the profiles of \lyg\ and \lyd. There are no metal
lines in the system.

{\it \zabs\ = 2.8493 ---} The central trough was fit with five
components referring to the profile of \lyd. Both sides of the \lyb\
line could not be fit well by our model. If additional components were
included in the fit, the model was found to absorb too much at \lya\,
as compared with the observed spectrum. No metal lines were detected
in our spectrum, though the corresponding \ion{C}{4} and \ion{Si}{4}
lines are not in the observed range.

{\it \zabs\ = 2.9105 ---} The observed spectrum could not be correctly
normalized, as this partial DLA system has strong damping wings. We
fit the strong trough with 8 components, using the profiles of lines
of higher order than \lyb. The column density of the main component is
\lognhi\ = 19.90, which is almost identical to the value determined by
Outram et al. (1999), \lognhi\ = 19.80.

\item{\it Q1937-1009 (\zem\ = 3.806). ---} LLS with \lognhi\ = 17.86
was detected at \zabs\ = 3.572 by Burles \& Tytler (1997). The LLS is
an ideal system for obtaining an estimate of the primordial value of
D/H, because the system has very low metallicity -- less than
$10^{-2}$ solar abundance. The D/H ratio has been evaluated to be D/H
= 3.3$\pm$0.3 $\times$ $10^{-5}$ at the 67~\% confidence level
(Tytler, Fan, \& Burles 1996; Burles \& Tytler 1998a). The LLS is
accompanied by various metal lines, such as \ion{C}{2}, \ion{C}{3},
\ion{C}{4}, \ion{N}{3}, \ion{Si}{2}, \ion{Si}{3}, \ion{Si}{4},
\ion{Fe}{2}, and \ion{Fe}{3}. Our spectrum ranges from 3890~\AA\ to
7450~\AA. Both \lya\ and \lyb\ are detected at \zabs\ = 2.792 --
3.806.

{\it \zabs\ = 3.5725 ---} Burles \& Tytler (1998a) found a candidate
\ion{D}{1} line in this system. They measured the column density of
the \hi\ line to be \lognhi\ = 17.86. We fit the central trough with
two components having column densities of \lognhi\ = 17.94 and
15.89. Various metal lines such as \ion{C}{2}, \ion{C}{3}, \ion{C}{4},
\ion{Si}{2}, \ion{Si}{3}, and \ion{Si}{4} were also detected in the
system, though all of them have only a single component.

\item{\it HS1946+7658 (\zem\ = 3.051). ---} This quasar was discovered
by Hagen et al. (1992) using objective prism observations at the 80 cm
Calar Alto Schmidt telescope. Three metal absorption systems have been
found. Two of them are highly ionized systems at \zabs\ = 3.049 and
2.843 with \ion{C}{4}, \ion{Si}{4}, and \ion{N}{5} lines. The other
one is a low ionization system at \zabs\ = 1.738 with \ion{Mg}{2} and
\ion{Fe}{2} lines (Hagen et al. 1992; Sadakane et al. 1993). Our
spectrum ranges from 3890~\AA\ to 6300~\AA. Both \lya\ and \lyb\ are
detected at \zabs\ = 2.792 -- 3.051. The system at \zabs\ = 2.843 is
located in the spectral gap.

{\it \zabs\ = 3.0498 ---} As this system is within 1000~\kms\ of the
emission redshift of the quasar, there are only a few absorption lines
redward of the main component. The detection of \ion{N}{5} lines in
this system strongly suggests that the system is highly ionized by the
UV flux of the quasar. This system is probably intrinsically
associated with the quasar itself, because Misawa et al. (2007) found
partial coverage in the \ion{N}{5} doublet. Other metal lines, such as
\ion{C}{4}, \ion{Si}{2}, \ion{Si}{3}, and \ion{Si}{4}, were also
detected.

\item{\it Q2223+2024 (\zem\ = 3.560). ---} This quasar was first
discovered by the MIT-Green Bank III 5 GHz survey (Griffith et
al. 1990). Storrie-Lombardi \& Wolfe (2000) confirmed that there was
no DLA candidate in their spectrum with a range of 2.101 $<$ $z$ $<$
3.514. No detections of absorption systems have been published for
this quasar. Our spectrum has a range of 4120~\AA\ to 6520~\AA. Both
\lya\ and \lyb\ are detected at \zabs\ = 3.017 -- 3.560.

\item{\it Q2344+2024 (\zem\ = 2.763). ---} This quasar is only $\sim$
5$^{\prime}$ separated on the sky from the quasar
Q2343+125. Q2344+2024 has \ion{C}{4} systems at \zabs\ = 2.427 and
2.429, while Q2343+125 has corresponding \ion{C}{4} systems at \zabs\
= 2.429 and 2.431 (SBS). The velocity separation of these systems
along the line of sight is less than 1000~\kms. These common
\ion{C}{4} systems could be produced by a Mpc-scale absorber, such as
a cluster of galaxies at the given redshift, although Misawa et
al. (2006) did not find any H$\alpha$ emitting galaxies at $z$ $\sim$
2.43 in this pair quasar field, down to $f$(H$\alpha$) of
1.6$\times$10$^{-17}$ erg~s$^{-1}$~cm$^{-2}$. Q2344+2024 has other
\ion{C}{4} systems at \zabs\ = 2.2754, 2.4265, 2.4371, 2.6964, 2.7017,
and 2.7817. Our spectrum has a range of 3410~\AA\ to 4940~\AA. Both
\lya\ and \lyb\ were detected at \zabs\ = 2.324 -- 2.763.

{\it \zabs\ = 2.4261 ---} The spectrum covers only \lya\ and \lyb\ in
this system. The model fit is poor, because both \lya\ and \lyb\ are
in regions with low S/N ratio (S/N = 16 at \lya\ and 6.8 at
\lyb). Weak \ion{Si}{4} doublet lines are also detected in this
system.

{\it \zabs\ = 2.6356 ---} We detected only a few \hi\ components in
the system; this is partially due to low S/N ratio. The Doppler
parameter of the main component, $b$ = 62~\kms, is larger than would
be expected for a \hi\ line, suggesting that the main component may be
separated into several narrow components. We did not detect any metal
lines in the system.

{\it \zabs\ = 2.7107 ---} The component that has the second largest
column density, \lognhi\ = 15.26, is separated by $-$750~\kms\ from
the main component, which has \lognhi\ = 16.64. It is unclear whether
the two components are physically related to each other. This system
is within 5000~\kms\ of the emission redshift of the quasar.

{\it \zabs\ = 2.7469 ---} The \hi\ component that has second largest
column density, \lognhi\ = 16.27, is redshifted $+$560~\kms\ from the
main component, which has \lognhi\ = 16.67. It is unclear whether the
two components are physically related. There are no corresponding
metal lines in the system. This system is within 1000~\kms\ of the
emission redshift of the quasar.

\end{description}


\clearpage


\begin{deluxetable}{rrrcccrrl}
\tabletypesize{\scriptsize}
\tablecaption{(Sample) Absorption lines in 86 \hi\ systems}
\tablewidth{0pt}
\tablehead{
\colhead{(1)} &
\colhead{(2)} &
\colhead{(3)} &
\colhead{(4)} &
\colhead{(5)} &
\colhead{(6)} &
\colhead{(7)} &
\colhead{(8)} &
\colhead{(9)} \\
\colhead{No.} &
\colhead{$\lambda_{obs}$} &
\colhead{$\Delta v$} &
\colhead{$z_{abs}$} &
\colhead{$\log N$} &
\colhead{$\sigma (\log N)$} &
\colhead{$b$} &
\colhead{$\sigma (b)$} &
\colhead{ID} \\
\colhead{} &
\colhead{(\AA)} &
\colhead{(\kms)} &
\colhead{} &
\colhead{(\cmm)} &
\colhead{(\cmm)} &
\colhead{(\kms)} &
\colhead{(\kms)} &
\colhead{} \\
}
\startdata
\hline
\multicolumn{9}{c}{Q0004+1711 (\zem=2.890)} \\
\hline
 \\
 \multicolumn{4}{c}{} &
 \multicolumn{2}{c}{\zabs=2.8284} &
 \multicolumn{3}{c}{} \\
 \cline{5-6}
 \\
1  & 4639.9 & -917.4 & 2.81671 & 13.505 & 0.018 & 29.48 & 1.43 & {\rm H}~{\sc i}~$\lambda$1216 \\
2  & 4641.1 & -837.6 & 2.81772 & 13.246 & 0.027 & 27.24 & 1.99 & {\rm H}~{\sc i}~$\lambda$1216 \\
3  & 4643.0 & -717.1 & 2.81926 & 12.739 & 0.069 & 11.86 & 1.81 & {\rm M}~{\sc i} \\
4  & 4644.0 & -650.4 & 2.82011 & 13.048 & 0.164 & 41.21 &19.08 & {\rm H}~{\sc i}~$\lambda$1216 \\
5  & 4645.8 & -531.6 & 2.82162 & 14.459 & 0.019 & 28.88 & 0.47 & {\rm H}~{\sc i}~$\lambda$1216 \\
6  & 4649.6 & -289.2 & 2.82471 & 13.244 & 0.040 & 31.27 & 3.82 & {\rm H}~{\sc i}~$\lambda$1216 \\
7  & 4654.1 &    0.0 & 2.82840 & 15.506 & 0.040 & 43.76 & 0.61 & {\rm H}~{\sc i}~$\lambda$1216 \\
8  & 4656.1 &  132.7 & 2.83009 & 13.087 & 0.057 & 25.87 & 3.96 & {\rm H}~{\sc i}~$\lambda$1216 \\
9  & 4656.6 &  161.0 & 2.83045 & 12.996 & 0.042 &  5.13 & 0.52 & {\rm M}~{\sc i} \\
10 & 4656.8 &  177.5 & 2.83066 & 13.103 & 0.035 &  8.31 & 0.88 & {\rm M}~{\sc i} \\
11 & 4657.2 &  203.4 & 2.83100 & 13.136 & 0.046 & 31.44 & 3.96 & {\rm H}~{\sc i}~$\lambda$1216 \\
12 & 4667.6 &  869.2 & 2.83951 & 13.722 & 0.014 & 73.98 & 3.36 & {\rm H}~{\sc i}~$\lambda$1216 \\
13 & 4669.6 &  999.2 & 2.84117 & 14.130 & 0.030 & 22.09 & 0.78 & {\rm H}~{\sc i}~$\lambda$1216 \\
\enddata
\end{deluxetable}


\begin{deluxetable}{rrrcccrrl}
\tabletypesize{\scriptsize}
\tablecaption{(Sample) Detected Metal Absorption lines in 86 \hi\ systems}
\tablewidth{0pt}
\tablehead{
\colhead{(1)} &
\colhead{(2)} &
\colhead{(3)} &
\colhead{(4)} &
\colhead{(5)} &
\colhead{(6)} &
\colhead{(7)} &
\colhead{(8)} &
\colhead{(9)} \\
\colhead{quasar} &
\colhead{$z_{sys}$} &
\colhead{$\Delta v$} &
\colhead{$z_{abs}$} &
\colhead{$\log N$} &
\colhead{$\sigma (\log N)$} &
\colhead{$b$} &
\colhead{$\sigma (b)$} &
\colhead{ID} \\
\colhead{} &
\colhead{} &
\colhead{(\kms)} &
\colhead{} &
\colhead{(\cmm)} &
\colhead{(\cmm)} &
\colhead{(\kms)} &
\colhead{(\kms)} &
\colhead{} \\
}
\startdata
Q0001+1711 & 2.8707 &  -26.0 & 2.87034 & 14.052 & 0.083 & 26.77 & 2.00 & {\rm C}~{\sc iii}  \\
           &        &  -23.9 & 2.87037 & 14.673 & 0.038 & 21.05 & 0.82 & {\rm C}~{\sc ii}   \\
           &        &  -23.7 & 2.87037 & 14.063 & 0.037 & 17.27 & 0.39 & {\rm Si}~{\sc ii}  \\
\hline
Q0014+8118 & 2.7989 &  -86.1 & 2.79782 & 12.634 & 0.015 & 21.72 & 0.93 & {\rm Si}~{\sc iii} \\
           &        &  -57.7 & 2.79818 & 12.387 & 0.021 &  6.60 & 0.26 & {\rm Si}~{\sc iii} \\
           &        &  -56.5 & 2.79820 & 12.950 & 0.016 &  6.96 & 0.27 & {\rm C}~{\sc iv}   \\
           &        &  -56.2 & 2.79820 & 12.537 & 0.010 &  6.91 & 0.17 & {\rm Si}~{\sc iv}  \\
           &        &   14.9 & 2.79910 & 13.032 & 0.024 & 18.89 & 1.14 & {\rm C}~{\sc iv}   \\
           &        &   17.5 & 2.79913 & 12.204 & 0.082 & 11.78 & 2.55 & {\rm Si}~{\sc iv}  \\
           &        &   34.4 & 2.79935 & 12.055 & 0.206 &  5.72 & 2.02 & {\rm C}~{\sc iv}   \\
           &        &   34.7 & 2.79935 & 11.456 & 0.312 &  5.19 & 2.23 & {\rm Si}~{\sc iv}  \\
           &        &   58.2 & 2.79965 & 12.536 & 0.046 & 14.24 & 1.05 & {\rm Si}~{\sc iv}  \\
           &        &   61.7 & 2.79969 & 13.525 & 0.015 & 32.73 & 1.20 & {\rm Si}~{\sc iii} \\
           &        &   67.3 & 2.79977 & 13.576 & 0.022 & 25.91 & 1.59 & {\rm C}~{\sc iv}   \\
\enddata
\end{deluxetable}



\clearpage


\begin{figure}
\centerline{
\includegraphics[width=4cm]{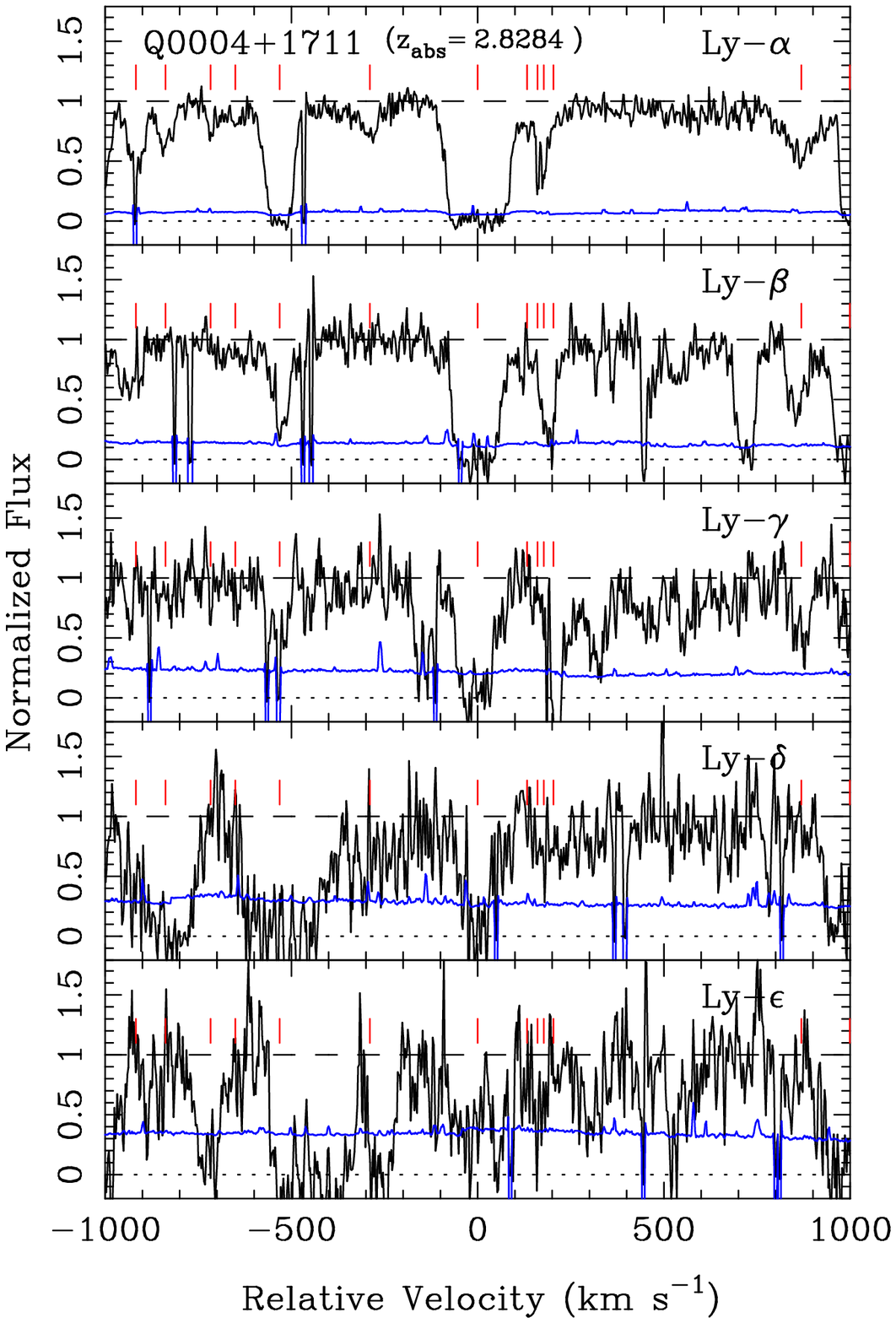}
}
\caption{(Sample) Velocity maps of the lowest five orders of Lyman
series (i.e., \lya, \lyb, \lyg, \lyd, and \lye) in a \hi\ system at
\zabs\ = 2.8284 in the spectrum of Q0004+1711. A histogram just above
the zero flux is 1$\sigma$ error spectrum. Tick marks above the flux
spectrum denote the positions of \ion{H}{1} and \ion{M}{1}
lines. Metal lines in the velocity window are not marked because they
are not physically related to the \hi\ system.}
\end{figure}


\begin{thebibliography}{}
\bibitem[Bahcall et al. 1992]{bah92} Bahcall, J.N., Hartig, G.F.,
  Jannuzi, B.T., Maoz, D., and Schneider, D.P., 1992, \apj, 400, L51
\bibitem[Bajtlik, Duncan, and Ostriker 1988]{baj88} Bajtlik, S.,
  Duncan, R.C., and Ostriker, J.P., 1988, \apj, 327, 570
\bibitem[Barthel, Tytler, and Thomson 1990]{bar90} Barthel, P.D.,
  Tytler, D.R., and Thomson, B., 1990, \aaps, 82, 339
\bibitem[Barvainis and Ivison 2002]{barv02} Barvainis, R.I. and
  Ivison, R., 2002, \apj, 571, 712
\bibitem[Bechtold 1994]{bec94} Bechtold, J., 1994, \apjs, 91, 1
\bibitem[Becker et al. 1991]{bec91} Becker, R.H., White, R.L., and
  Edwards, A.L., 1991, \apjs, 75,1
\bibitem[Bechtold 1994]{bec94} Bechtold, J., 1994, \apjs, 91, 1
\bibitem[Bergeron and Boiss\'e 1991]{ber91} Bergeron, J., and
  Boiss\'e, P., 1991, \aap, 243, 344
\bibitem[Burles, Kirkman, and Tytler 1999]{bur99} Burles, S., Kirkman,
  D., and Tytler, D., 1999, \apj, 519, 18
\bibitem[Burles and Tytler 1998a]{bur98a} Burles, S., and Tytler, D.,
  1998a, \apj, 499, 699
\bibitem[Burles and Tytler 1998b]{bur98b} Burles, S., and Tytler, D.,
  1998b, \apj, 507, 732
\bibitem[Burles and Tytler 1997]{bur97b} Burles, S., and Tytler, D.,
  1997, \aj, 114, 1330
\bibitem[Carballo et al. 1995]{car95} Carballo, R., Barcons, X., and
  Webb, J.K., 1995, \aj, 109, 1531
\bibitem[Carswell 1996]{car96} Carswell, R.F., et al., 1996, \mnras,
  278, 506
\bibitem[Carswell et al. 1994]{car94} Carswell, R.F., Rauch, M.,
  Weymann, R.J., Cooke, A.J., and Webb, J.K., 1994, \mnras, 268, L1
\bibitem[Carswell et al. 1984]{car84} Carswell, R.~F., {Morton},
  D.~C., {Smith}, M.~G., {Stockton}, A.~N., {Turnshek}, D.~A. and
  {Weymann}, R.~J.  1984 {\apj}, 278, 486
\bibitem[Carswell et al. 1982]{car82} Carswell, R.F., Whelan, J.A.J.,
  Smith, M.G., Boksenberg, A., and Tytler, D., 1982, \mnras, 198, 91
\bibitem[Carswell et al. 1975]{car75} Carswell, R.F., Strittmatter,
  P.A., Williams, R.E., Beaver, E.A., and Harms, R., 1975, \apj, 195,
  269
\bibitem[Chen, Lanzetta, and Webb 2001]{che01} Chen, H.-W., Lanzetta,
  K.M., and Webb, J.K., 2001, \apj, 556, 158
\bibitem[Chen et al. 1981]{che81} Chen, H.-W., Morton, D.C., Peterson,
  B.A., Wright, A.E., and Jauncey, D.L., 1981, \mnras, 196, 715
\bibitem[Crampton et al. 1989]{cra89} Crampton, D., Cowley, A.P., and
  Hartwick, F.D.A., 1989, \apj, 345, 59
\bibitem[Crampton et al. 1988]{cra88} Crampton, D., Cowley, A.P.,
  Schmidtke, P., Janson, T., and Durrell, P., 1988, \aj, 96, 816
\bibitem[Cristiani et al. 1997]{cri97} Cristiani, S., D'Odorico, S.,
  D'Odorico, V., Fontana, A., Giallongo, E., and Savaglio, S., 1997,
  \mnras, 285, 209
\bibitem[Dave and Tripp 2001]{dav01} Dav{\'e}, R., \& Tripp, T.~M.,
  2001, \apj, 553, 528
\bibitem[Dave et al. 1999]{dav99} Dav\'e, R., Hernquist, L., Katz, N.,
  and Weinberg, D.H., 1999, \apj, 511, 521
\bibitem[Dobrzycki et al. 1996]{dob96} Dobrzycki, A., Engels, D.,
  Hagen, H.-J., Elvis, M., Huchra, J., and Reimers, D., 1996, \baas,
  188.0602
\bibitem[Griffith et al. 1990]{gri90} Griffith, M., Langston, G.,
  Heflin, M., Conner, S., Lehar, J., and Burke, B., 1990, \apjs, 74,
  128
\bibitem[Hagen et al. 1995]{hag95} Hagen, H.-J., Groote, D., Engels,
  D., and Reimers, D., 1995, \aaps, 111, 195
\bibitem[Hagen et al. 1992]{hag92} Hagen, H.-J., Cordis, L., Engels,
  D., Groote, D., Haug, U., Heber, U., K$\ddot{o}$hler, T., Wisotzki,
  L., and Reimers, D., 1992, \aap, 253, L5
\bibitem[Hewitt and Burbidge 1987]{hew87} Hewitt, A., and Burbidge,
  G., 1987, \apjs, 63,1
\bibitem[Hu et al. 1995]{hu95} Hu, E., Kim, T.-S., Cowie, L.L.,
  Songaila, A., and Rauch, M., 1995, \aj, 110, 1526 (H95)
\bibitem[Janknecht et al. 2006]{jan06} Janknecht, E., Reimers, D.,
  Lopez, S., and Tytler, D., 2006, \aap, 458, 427
\bibitem[Kim et al. 2002a]{kim02a} Kim, T.-S., Cristiani, S., and
  D'Odorico, S., 2002a, \aap, 383, 747
\bibitem[Kim et al. 2002b]{kim02b} Kim, T.-S., Carswell, R.F.,
  Cristiani, S., D'Odorico, S., and Giallongo, E., 2002b, \mnras, 335,
  555
\bibitem[Kim et al. 2001]{kim01} Kim, T.-S., Cristiani, S., and
  D'Odorico, S., 2001, \aap, 373, 757
\bibitem[Kirkman et al. 2005]{kir05} Kirkman, D., Tytler, D., Suzuki,
  N., Melis, C., Hollywood, S., James, K., So, G., Lubin, D., Jena,
  T., Norman, M.L., and Paschos, P., 2005, \mnras, 360, 1373
\bibitem[Kirkman et al. 2003]{kir03} Kirkman, D., Tytler, D., Suzuki,
  N., O'Meara, J.M., and Lubin, D., 2003, \apjs, 149, 1
\bibitem[Kirkman et al. 2000]{kir00} Kirkman, D., Tytlrt, D., Burles,
  S., Lubin, D., and O'Meara, J.M., 2000, \apj, 529, 655
\bibitem[Kirkman and Tytler 1999]{kir99} Kirkman, D., and Tytler, D.,
  1999, \apj, 512, L5
\bibitem[Kirkman and Tytler 1997a]{kir97a} Kirkman, D., and Tytler, D.,
  1997a, \apj, 484, 672 (KT97)
\bibitem[Kirkman and Tytler 1997b]{kir97b} Kirkman, D., and Tytler, D.,
  1997b, \apj, 489, L123
\bibitem[Kohler et al. 1999]{koh99} K$\ddot{o}$hler, S., Reimers, D.,
  Tytler, D., Hagen, H.-J., Barlow, T., and Burles, S., 1999, \aap,
  342, 395
\bibitem[Kormann et al. 1994]{kor94} Kormann, R., Schneider, P., and
  Bartelmann, M., 1994, \aap, 286, 357
\bibitem[Kuhr et al. 1983]{kuh83} Kuhr, H., Liebert, J.W.,
  Strittmater, P.A., Schmidt, G.D., and Mackay, C., 1983, \apj, 275,
  L33
\bibitem[Lanzetta et al. 1991]{lan91b} Lanzetta, K.M., Wolfe, A.M.,
  Turnshek, D.A., Lu, L., McMahon, R.G., and Hazard, C., 1991, \apjs,
  77, 1
\bibitem[Lanzetta 1991]{lan91a} Lanzetta, K.M., 1991, \apj, 375, 1
\bibitem[Lehner et al. 2007 ]{leh07} Lehner, N., Savage, B. D.,
  Richter, P., Sembach, K. R., Tripp, T. M., \& Wakker, B. P., 2007,
  \apj, 658, 680
\bibitem[Lu, Sargent, and Barlow 1998]{lu98} Lu, L., Sargent, W.L.W.,
  and Barlow, T.A., 1998, \aj, 115, 55
\bibitem[Lu et al. 1996a]{lu96a} Lu, L., Sargent, W.L.W., Womble, D.S.,
  and Takada-Hidai, M., 1996a, \apj, 472, 509 (L96)
\bibitem[Lu et al. 1996b]{lu96b} Lu, L., Wallace, L., Sargent, W., and
  Barlow, T.A., 1996b, \apjs, 107, 475
\bibitem[Lu et al. 1993]{lu93} Lu, L., Wolfe, A.M., Turnshek, D.A.,
  and Lanzetta, K.M., 1993, \apjs, 84, 1
\bibitem[Melott 1980]{mel80} Melott, A.L., 1980, \apj, 241, 889
\bibitem[Misawa et al. 2007]{mis07} Misawa, T., Charlton, J.C.,
  Eracleous, M., Ganguly, R., Tytler, D., Kirkman, D., Suzuki, N., and
  Lubin, D., 2007, \apjs, in press, astro-ph/0702101
\bibitem[Misawa et al. 2006]{mis06} Misawa, T., Kashikawa, N.,
  Ohyama, Y., Hashimoto, T., and Iye, M., 2006, \aj, 131, 34
\bibitem[Misawa et al. 2004]{mis04} Misawa, T., Tytler, D., Iye, M.,
  Paschos, P., Norman, M., Kirkman, D., O'Meara, J., Suzuki, N.,
  Kashikawa, N., 2004, \aj, 128, 2954
\bibitem[Misawa et al. 2002]{mis02} Misawa, T., Tytler, D., Iye, M.,
  Storrie-Lombardi, L.J., Suzuki, N., and Wolfe, A.M., 2002, \aj, 123,
  1847
\bibitem[Monet et al. 2003]{mon03} Monet, D.G., et al. 2003, \aj, 125,
  984
\bibitem[Monet et al. 1998]{mon98} Monet, D. et al., 1998, USNO-A2.0
  (Flagstaff: US Nav. Obs.)
\bibitem[Murdoch et al. 1986]{mur86} Murdoch, H.S., Hunstead, R.W.,
  Pettini, M., and Blades, J.C., 1986, \apj, 309, 19
\bibitem[O'Meara et al. 2001]{ome01} O'Meara, J.M., Tytler, D.,
  Kirkman, D., Suzuki, N., Prochaska, J.X., Lubin, D., and Wolfe,
  A.M., 2001, \apj, 552, 718
\bibitem[Osmer and Smith 1976]{osm76} Osmer, P.S., and Smith, M.G.,
  1976, \apj, 210, 267
\bibitem[Outram et al. 1999]{out99} Outram, P.J., Chaffee, F.H., and
  Carswell, R.F., 1999, \mnras, 310, 289
\bibitem[Penton et al. 2004]{pen04} Penton, S. V., Stocke, J. T., \&
  Shull, J. M., 2004, \apjs, 152, 29
\bibitem[Penton et al. 2002]{pen02} Penton, S. V., Stocke, J. T., \&
  Shull, J. M., 2002, \apj, 565, 720
\bibitem[Peroux et al. 2001]{per01} P\'eroux, C., Storrie-Lombardi,
  L.J., McMahon, R.G., Irwin, M., and Hook, I.M., 2001, \apj, 121,
  1799
\bibitem[Petitjean, Rauch, and Carswell 1994]{pet94} Petitjean, P.,
  Rauch, M., and Carswell, R.F., 1994, \aap, 291, 29
\bibitem[Petitjean et al. 1993]{pet93} Petitjean, P., Webb, J.K.,
  Rauch, M., Carswell, R.F., and Lanzetta, K., 1993, \mnras, 262, 499
\bibitem[Prochaska et al. 2001]{pro01} Prochaska, J.X., Wolfe, A.M.,
  Tytler, D., Burles, S., Cooke, J., Gawiser, E., Kirkman, D.,
  O'Meara, J.M., and Storrie-Lombardi, L., 2001, \apjs, 137, 21
\bibitem[Rauch et al. 1998]{rau98} Rauch, M., 1998, ARA\&A, 36, 267
\bibitem[Rauch et al. 1992]{rau92} Rauch, M., Carswell, R.F., Chaffee,
  F.H., Foltz, C.B., Webb, J.K., Weymann, R.J., Bechtold, J., and
  Green, R.F., 1992, \apj, 390, 387
\bibitem[Reimers et al. 1995]{rei95} Reimers, D., Rodriguez-Pascual,
  P., Hagen, H.-J., and Wisotzki, L., 1995, \aap, 293, L21
\bibitem[Reimers et al. 1992]{rei92} Reimers, D., Vogel, S., Hagen,
  H.-J., Engels, D., Groote, D., Wamsteker, W., Clavel, J., and Rosa,
  M.R., 1992, \nat, 360, 561
\bibitem[Reimers et al. 1989]{rei89} Reimers, D., Clavel, J., Groote,
  D., Engels, D., Hagen, H.-J., Naylor, T., Wamsteker, W., and Hopp,
  U., 1989, \aap, 218, 71
\bibitem[Rodriguez-Pascual et al. 1995]{rod95} Rodr\'iguez-Pascual,
  P.M., Fuente, A., Sanz, J.L., Recondo, M.C., Clavel, J.,
  Santos-Lle\'o, M., and Wamsteker, W., 1995, \apj, 448, 575
\bibitem[Rugers and Hogan 1996a]{rug96a} Rugers, M., and Hogan, C.J.,
  1996a, \apj, 459, L1
\bibitem[Rugers and Hogan 1996b]{rug96b} Rugers, M., and Hogan, C.J.,
  1996b, \aj, 111, 2135
\bibitem[Sadakane et al. 1993]{sad93} Sadakane, K., Takada-Hidat, M.,
  Yoshida, M., Kosugi, G., and Ohtani, H., 1993, \pasj, 45, 505
\bibitem[Sanz et al. 1993]{san93} Sanz, J.L., Clavel, J., Naylor, T.,
  and Wamsteker, W., 1993, \mnras, 260, 468
\bibitem[Sargent, Steidel, and Boksenberg 1989]{sar89} Sargent,
  W.L.W., Steidel, C.C., and Boksenberg, A., 1989, \apj, 69, 703 (SSB)
\bibitem[Sargent, Boksenberg, and Steidel 1988]{sar88} Sargent,
  W.L.W., Boksenberg, A., and Steidel, C.C., 1988, \apjs, 68, 539
  (SBS)
\bibitem[Sargent et al. 1980]{sar80} Sargent, W.L.W., Young, P.J.,
  Boksenberg, A., and Tytler, D., 1980, \apjs, 42, 41
\bibitem[Schneider, Schmidt, and Gunn 1994]{sch94} Schneider, D.P.,
  Schmidt, M., and Gunn, J.E., 1994, \aj, 107, 1245
\bibitem[Songaila, Wampler, and Cowie]{son97} Songaila, A., Wampler,
  E.J., and Cowie, L.L., 1997, \nat, 385, 137
\bibitem[Songaila and Cowie 1996]{son96} Songaila, A., and Cowie,
  L.L., 1996, \aj, 112, 335
\bibitem[Songaila et al. 1994]{son94} Songaila, A., Cowie, L.L.,
  Hogan, C.J., and Rugers, M., 1994, \nat, 368, 599
\bibitem[Steidel and Sargent 1992]{ste92} Steidel, C.C. and Sargetnt,
  W.L.W., 1992, \apjs, 80, 1
\bibitem[Steidel 1990a]{ste90a} Steidel, C.C., 1990a, \apjs, 74, 37
\bibitem[Steidel 1990b]{ste90b} Steidel, C.C., 1990b, \apjs, 72, 1
\bibitem[Stengler-Larrea et al. 1995]{ste95} Stengler-Larrea, E.A.,
  Boksenberg, A., Steidel, C.C., Sargent, W.L.W., Bahcall, J.N.,
  Bergeron, J., Hartig, G.F., Jannuzi, B.T., Kirhakos, S., Savage,
  B.D., Schneider, D.P., Turnshek, D.A., and Weymann, R.J., 1995,
  \apj, 444, 64
\bibitem[Stepanian et al. 1996]{ste96} Stepanian, J.A., Chavushian,
  V.H., Chaffee, F.H., Foltz, C.B., and Green, R.F., 1996, \aap, 309,
  702
\bibitem[Stepanian et al. 1990]{ste90} Stepanian, J.A., Lipovetsky,
  V.A., and Erastova, 1990, Astrophyzica, 32, 441
\bibitem[Storrie-Lombardi and Wolfe 2000]{sto00} Storrie-Lombardi,
  L.J., and Wolfe, A.M., 2000, \apj, 543, 552
\bibitem[Storrie-Lombardi et al. 1996]{sto96} Storrie-Lombardi, L.J.,
  McMahon, R.G., Irwin, M.J., and Hazard, C., 1996, \apj, 468, 121
\bibitem[Theuns et al. 1998]{the98} Theuns, T., Leonard, A., \&
  Efstathiou, G., 1998, \mnras, 297, L49
\bibitem[Tytler, Fa, and Burles 1996]{tyt96} Tytler, D., Fan, X.-M.,
  and Burles, S., 1996, \nat, 381, 207
\bibitem[Tytler 1987]{tyt87} Tytler, D., 1987, \apj, 321, 49
\bibitem[Tytler 1982]{tyt82} Tytler, D., 1982, \nat, 298, 427
\bibitem[Veron-Cetty and Veron 2003]{ver03} V\'eron-Cetty, M.-P. and
  V\'eron, P., 2003, \aap, 412, 399
\bibitem[Wampler et al. 1996]{wam96} Wampler, E.J., Williger, G.M.,
  Baldwin, J.A., Carswell, R.F., Hazard, C., and McMahon, R.G., 1996,
  \aap, 316, 33
\bibitem[Webb 1987]{web87} Webb, J.K., 1987, in IAU Symp. 124,
  Observational Cosmology, ed. A.Hewitt, G.Burbidge, and L.Z.Fang
  (Dordrecht:Reidel), 803
\bibitem[Weymann et al. 1998a]{wey98a} Weymann, R.J., Jannuzi, B.T.,
  Lu, L., Bahcall, J.N., Bergeron, J., Boksenberg, A., Hartig, G.F.,
  Kirhakos, S., Sargent, W.L.W., Savage, B.D., Schneider, D.P.,
  Turnshek, D.A., and Wolfe, A.M., 1998a, \apj, 506, 1
\bibitem[Weymann et al. 1998b]{wey98b} Weymann, R. J., et al., 1998b,
  \apj, 506, 1
\bibitem[Wolfe et al. 1995]{wol95} Wolfe, A.M., Lanzetta, K.M., Foltz,
  C.B., and Chaffee, F.H., 1995, \apj, 454, 698
\bibitem[Young et al. 1979]{you79} Young, P.J., Sargent, W.L.W.,
  Boksenberg, A., Carswell, R.F., and Whelan, J.A.J., 1979, \apj, 229,
  891
\bibitem[Zhang et al. 1997]{zha97} Zhang, Y., Anninos, P., Norman,
  M.L., and Meiksin, A., 1997, \apj, 485, 496
\end{thebibliography}
\end{document}